\documentclass[reprint, nofootinbib,superscriptaddress]{revtex4-1}

\usepackage[toc,page]{appendix}
\usepackage[british]{babel}
\usepackage[utf8]{inputenc}
\usepackage{amsmath,amssymb}
\usepackage{graphicx}
\usepackage{ae}
\usepackage{esdiff}
\usepackage{braket}
\usepackage{bbm}
\usepackage{simplewick}
\usepackage{float}
\usepackage{braket}
\usepackage{simplewick}
\usepackage{color}
\usepackage{framed}
\usepackage[caption=false]{subfig}
\usepackage[pdftex,colorlinks=true,linkcolor=blue,citecolor=blue,urlcolor=black]{hyperref}
\usepackage{mathtools}
\usepackage{enumitem}
\usepackage{ragged2e}
\usepackage{floatflt}
\usepackage{import}
\usepackage{titlesec}
\usepackage{multirow}
\usepackage{etoolbox}
\usepackage{appendix}
\usepackage{dsfont}
\usepackage{comment}
\setlist[itemize]{leftmargin=*}

\makeatletter
\patchcmd{\ttlh@hang}{\parindent\z@}{\parindent\z@\leavevmode}{}{}
\patchcmd{\ttlh@hang}{\noindent}{}{}{}
\makeatother

\begin{document}

\newcommand{\norm}[1]{\left\lVert#1\right\rVert}
\newcommand{\fpch}[1]{{\color{blue}#1}}
\newcommand{\fpcm}[1]{{\color{red}FP: #1}}

\newcommand{\red}[1]{{\color{red}{#1}}}

\def \r{{\textbf{r}}}
\def \k{{\textbf{k}}}
\def \p{{\textbf{p}}}
\def \q{{\textbf{q}}}
\def \x{{\textbf{x}}}
\def \A{{\textbf{{A}}}}
\def \a{{\textbf{{a}}}}
\def \b{{\textbf{{b}}}}
\def \c{{\textbf{{c}}}}
\def \z{{\textbf{{z}}}}
\def \0{{\textbf{{0}}}}
\def \dl{\frac{\partial}{\partial l}}
\def \P{{\boldsymbol{P}}}
\def \K{{\boldsymbol{K}}}

\def\BigColSep{\setlength{\arraycolsep}{50pt}}

\title{ Non-Fermi liquid at the FFLO  quantum critical point}

\author{Dimitri Pimenov}\email{D.Pimenov@physik.lmu.de}
\affiliation{Physics Department, Arnold Sommerfeld Center for Theoretical Physics, and Center for NanoScience, Ludwig-Maximilians-University Munich, 80333 Munich, Germany}

\author{Ipsita Mandal} 
\affiliation{Max-Planck Institute for the Physics of Complex Systems,
01187 Dresden, Germany}

\author{Francesco Piazza} 
\affiliation{Max-Planck Institute for the Physics of Complex Systems,
01187 Dresden, Germany}

\author{Matthias Punk}
\affiliation{Physics Department, Arnold Sommerfeld Center for Theoretical Physics, and Center for NanoScience, Ludwig-Maximilians-University Munich, 80333 Munich, Germany}

\begin{abstract}
When a 2D superconductor is subjected to a strong in-plane magnetic field, Zeeman polarization of the Fermi surface can give rise to inhomogeneous FFLO order with a spatially modulated gap. Further increase of the magnetic field eventually drives the system into a normal metal state. Here, we perform a renormalization group analysis of this quantum phase transition, starting from an appropriate low-energy theory recently introduced in Ref.\ \cite{piazza2016fflo}. We compute one-loop flow equations within the controlled dimensional regularization scheme with fixed dimension of Fermi surface, expanding in $\epsilon = 5/2 - d$. We find a new stable non-Fermi liquid fixed point and discuss its critical properties. One of the most interesting aspects of the FFLO non-Fermi liquid scenario is that the quantum critical point is potentially naked, with the scaling regime observable down to arbitrary low temperatures. In order to study this possibility, we perform a general analysis of competing instabilities, which suggests that only charge density wave order is enhanced in the vicinity of the quantum critical point.  
\end{abstract}
\maketitle

\section{Introduction}

A variety of strongly correlated electron materials show unusual metallic behavior, which cannot be described within Landau’s Fermi liquid theory. In many cases this non-Fermi liquid regime seems to be tied to the presence of a quantum critical point (QCP) between a normal metal and a different symmetry broken phase \cite{RevModPhys.79.1015}. One paradigmatic example are certain heavy Fermion materials, where the non-Fermi liquid regime seems to extend out of a QCP related to the onset of antiferromagnetic order \cite{RevModPhys.73.797}.

Of special interest and practical relevance are quasi two-dimensional systems, where the coupling between electrons and order parameter fluctuations in the vicinity of the QCP is particularly strong. This leads to a loss of electronic quasiparticle coherence due to an intricate interplay between electronic degrees of freedom and the order-parameter dynamics \cite{PhysRevB.50.14048, PhysRevB.50.17917, PhysRevLett.91.066402, PhysRevB.80.165102, metlitski2010quantum, Metlitski2010}. The fact that no well-defined quasiparticle excitations exist in such strongly coupled systems makes the theoretical description of these non-Fermi liquids especially challenging. 

Two notable theoretical developments added considerably to our understanding of such non-Fermi liquids. First, it was realized that models of fermions coupled to order parameter fluctuations can be numerically simulated using Quantum Monte Carlo techniques avoiding the infamous sign-problem under certain conditions \cite{Berg1606}. Second, it was shown that field-theoretical approaches can be controlled by increasing the co-dimension of the Fermi surface, which allows for the computation of critical exponents in a systematic epsilon expansion \cite{senthil2009fermi,dalidovich2013perturbative}. In this work we will make use of the latter ideas in particular.

\begin{figure}
\centering
\includegraphics[width=.7\columnwidth]{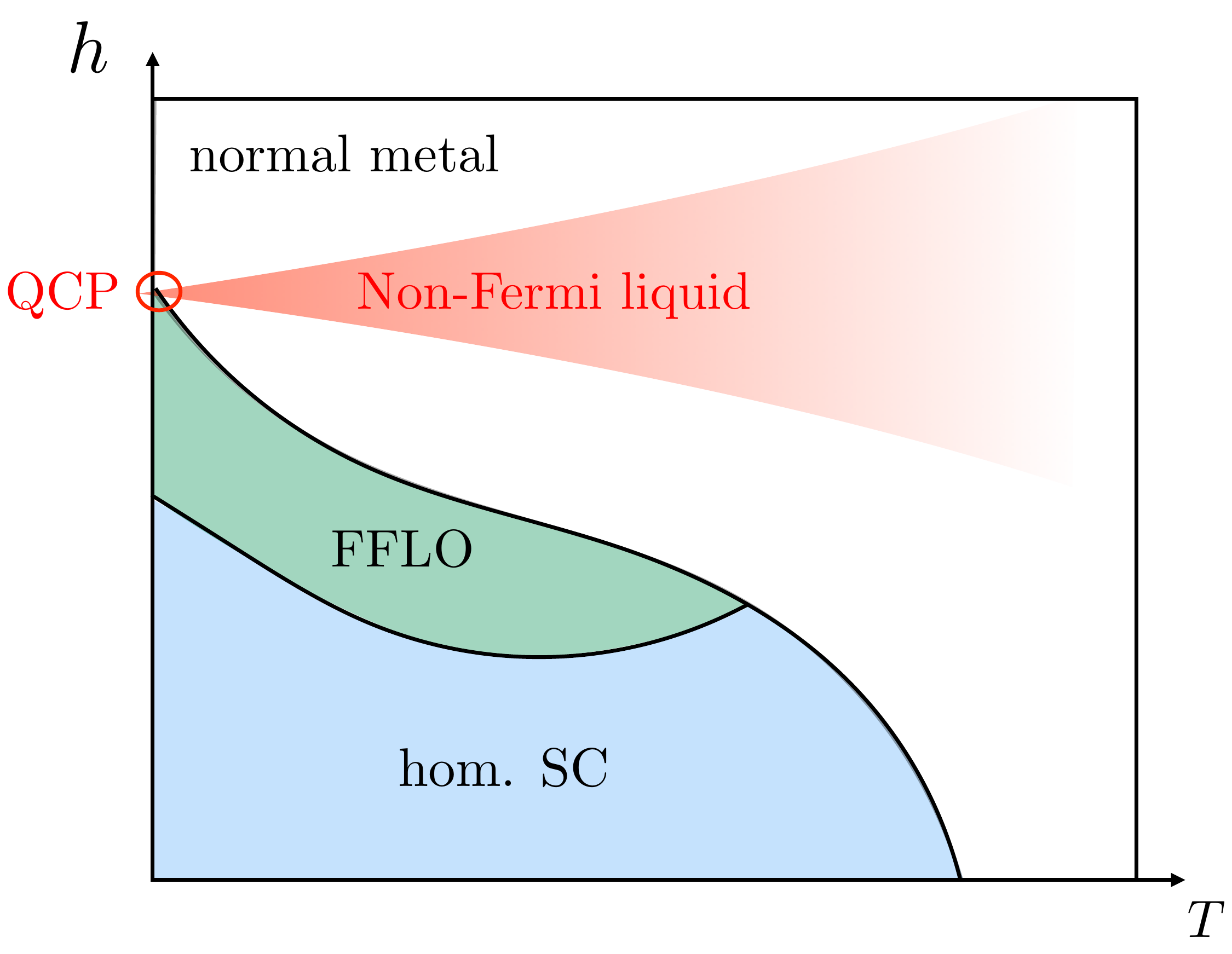}
\caption{(Color online) Typical temperature-magnetic field phase diagram of a superconductor susceptible to FFLO pairing. This picture was adapted from Ref.\ \cite{piazza2016fflo}.}
\label{phasediag}
\end{figure}

So far, most of the theoretical works focused on the experimentally relevant cases of spin-density wave or Ising-nematic critical points in metals. Here we consider a different problem instead and study the quantum critical point between a normal metal and an inhomogeneous Fulde-Ferell-Larkin-Ovchinnikov (FFLO) superconductor \cite{fulde1964superconductivity, Larkin1965} in two dimensions. This scenario was put forward by Piazza et al.\  \cite{piazza2016fflo}, who showed that, for appreciable in-plane anisotropy of the Fermi surface,  there is a strong coupling between electrons and FFLO fluctuations in the vicinity of hot spots on the Fermi surface, potentially giving rise to non-Fermi liquid behavior in the quantum critical regime extending from the QCP at finite temperature, see Fig.\ \ref{phasediag}. A similar treatment of the isotropic case can be found in Ref.\  \cite{samokhin2006quantum}.

The stabilization of FFLO phases requires clean superconducting materials with suppressed orbital pair breaking effects plus highly anisotropic Fermi surfaces, such as the ones shown by layered materials \cite{croitoru2017search}. Several strong indications of such phases are found in an increasing number of experimental cases, involving organic superconductors \cite{bechg_fflo,Jerome_fflo,mayaffre2014evidence,wosnitza_review}, heavy-fermion systems \cite{matsuda2007fulde,ptok2017influence}, iron-based superconductors\cite{iron_based_fflo1,iron_based_fflo2}, Al films \cite{films_fflo} as well as superconductor-ferromagnet bilayers \cite{bilayers_fflo_Aarts,bilayers_fflo_Buzdin}.

While the previous study \cite{piazza2016fflo} of FFLO non-Fermi liquid criticality was based on a perturbative, RPA-type approach, we will employ the epsilon expansion by Dalidovich and Lee \cite{dalidovich2013perturbative} in this work. This allows us to compute critical exponents in a systematic expansion around $d=5/2$ dimensions, similar to the Ising-nematic problem. 

One intriguing aspect of non-Fermi liquids in the vicinity of FFLO critical points is that the QCP is potentially „naked“ and not masked by a competing order. Indeed, in the Ising-nematic as well as the SDW scenarios, the order parameter fluctuations give rise to an effective attraction between the electrons, burying the QCP deep underneath a superconducting phase \cite{PhysRevB.34.6554, PhysRevB.34.8190, SachdevMetlitskiPunk, lederer2015enhancement, metlitski2014non, mandal2016superconducting, schattner2016competing, lederer2017superconductivity}. 
One consequence of this competing superconductivity is that the scaling regime of the QCP might be hardly accessible in experiments. By contrast, there is no obvious superconducting order parameter with a different symmetry competing with FFLO superconductivity, which could potentially mask the FFLO QCP. It might be possible, however, that other types of competing orders, such as charge density waves, are enhanced by fluctuations of the FFLO order parameter. We will discuss this issue in detail later in this work.

The rest of this paper is outlined as follows: First, we will give a non-technical overview of our main results and their physical consequences in Sec.\ \ref{summarysec}. Detailed computations are presented in the subsequent sections.  In Sec.\ \ref{crittheorysec}, the system under consideration is introduced, studied on mean field level, and lifted to higher dimensions. In Sec.\ \ref{diagsec}, we discuss one-loop quantum corrections, from which the renormalization group flow and critical properties are derived in Sec.\ \ref{rensec}. Possible competing instabilities are analyzed in Sec.\ \ref{compordsec}. Finally, a conclusion is presented in Sec.\ \ref{concsec}. Technical details of the computations are carried out in the Appendices. 

\section{Summary of results} 
\label{summarysec}

An appropriate field-theoretical description of the FFLO-normal metal quantum phase transition has to include dynamics of a bosonic FFLO order parameter $\Delta$ (a spatially modulated gap) coupled to the relevant ``slow'' electronic degrees of freedom $\psi$. As we show in Sec.\ \ref{crittheorysec}, such a description is accomplished by a low-energy action which contains 3 parameters $\{m, g, \delta v\}$. Here, $m$ is the ``boson mass'' resp.\ inverse correlation length, which is proportional to the deviation from the critical magnetic field $h_c$ and allows us to tune through the phase transition, $g$ is the strength of the electron-boson coupling (which is proportional to the microscopic electron attractive interaction) and $\delta v$ is a parameter which describes the relative spin-velocities of the electrons perpendicular to the Fermi surface (which we call the $k_x$-direction).

An RG analysis of this low-energy action which treats fermions and bosons on equal footing is the only rigorous way to 
 gain insight into the critical features of the transition, see e.g.\ chapter 18 of Ref.\ \cite{sachdev2011quantum} for an introduction. In the RG, the parameters of the low-energy action will flow as a function of the energy/length scale. In this work,  we  study the simplified flow of the interaction parameter $g$ at the quantum critical point ($m=0$), and also set  $\delta v = 0$ for technical reasons. 

The first goal of the RG analysis is to locate a fixed point $g = g^\star$, which gives access to critical exponents and correlations. To our knowledge, this was not yet accomplished in the study of FFLO criticality. Using an epsilon-expansion method introduced in the context of metallic quantum critical points \cite{dalidovich2013perturbative}, we find a stable fixed point corresponding to a continuous transition at $g^\star \propto \epsilon^{3/4}$, where $\epsilon = 5/2 -d = 1/2$.

\begin{table}
\begin{ruledtabular}
\begin{center}
    \begin{tabular}{lll}
       \multicolumn{1}{l}{\textbf{Critical Exponent}}  &  & \multicolumn{1}{l}{\textbf{Value in $d=2$ at $\mathcal{O}(\epsilon)$}} 
   \\ 
 	  \hline
 	\multicolumn{1}{ l }{$z$}&  \hspace{-5em} dyn.\ crit.\ exponent
     & $\phantom{-}3/2$  \\ 
    % \hline
     \multicolumn{1}{ l }{$\eta_\psi = \eta_{\Delta}$} & \hspace{-5em} anomalous dim.
     & $-1/4$ \\ 
  %   \hline
     \multicolumn{1}{ l }{$\nu$} & \hspace{-5em} corr.\ length.\ exp.\ 
    & $  \phantom {-} 1$ \\
    \end{tabular}
    \caption{Critical exponents at the FFLO fixed point, $g=g^\star$. Here, $z$ is the dynamical critical exponent, $\eta_\psi = \eta_\Delta$ are the anomalous dimension of fermions and bosons (which coincide in $\mathcal{O}(\epsilon)$, and $\nu$ is the correlation length exponent.}
    \label{tab0}    
\end{center}
\end{ruledtabular}
\end{table}

The critical exponents obtained in our analysis of this new fixed point are presented in Tab.\ \ref{tab0}.
In this table, $z$ is the dynamical critical exponent, which determines how the time-like direction scales compared to the space-like directions. $\eta_\psi, \eta_{\Delta}$ are the anomalous dimensions of the fermions and bosons (which coincide at one-loop level), i.e. the deviation from the scaling determined by power counting for the free theory. $\nu$ is the correlation length exponent, given by the inverse RG eigenvalue of the mass term $m$. 

The main value of these critical exponents lies in the fact that they determine the critical correlations, i.e. the electron and boson propagators. In accordance with the RPA-type treatment of Ref.\ \cite{piazza2016fflo} (which is thereby set on solid ground), the scaling forms of the two-point correlators in 2D agree with
\begin{align}
\label{twopointscalingelectrons}
& G(\omega, k_x,k_y) = \frac{1}{i\omega - \delta_k- \Sigma(\omega)},\\ & \delta_k \propto k_x + k_y^2, \ \text{Im}[\Sigma(\omega)] \propto g^{4/3} \omega^{2/3} \notag 
\end{align} 
for electrons. For bosons one obtains 
\begin{align}
\label{twopointscalingbosonssum}
& D(\omega, k_x,k_y) = \frac{1}{k_y^2 - \Pi(\omega,k_y)}, \ \Pi(\omega,k_y) \propto -g^{2} \frac{|\omega|}{|k_y|},   
\end{align} 
where $\Pi$ is the inverse pair propagator. The $k_x$-dependence of the boson propagator is irrelevant in the RG sense. The non-analytic behaviour of the self-energies supports our claim that the quantum critical point is of non-Fermi liquid type. 
Under assumption of $\omega/{T}$-scaling, signatures of these critical correlations are measurable in the non-Fermi liquid region indicated in Fig.\ \ref{phasediag}. This region is delimited by the two crossover lines satisfying $k_BT \sim |h-h_{c}|^{z\nu}$ with $z\nu=3/2$ according to our results. Examples for physical observables include:

\begin{itemize}
\item [-]
 \textit{magnetic susceptibility} $\chi$: a simple computation (see App.\ \ref{suscapp} and Ref.\ \cite{larkin2008fluctuation,samokhin2006quantum} ) shows that the fluctuation contribution to the magnetic susceptibility $\chi_\Delta$ scales as $\chi_\Delta \propto \log(h-h_c)$ .
\item[-] \textit{fermionic decay rate $\Gamma$ and density of states $\rho(\omega)$}: from (\ref{twopointscalingelectrons}) one immediately sees that the quasiparticle decay rate has a non-Fermi liquid-like power law dependence $\Gamma \propto \omega^{2/3}$, while by integrating the spectral function over momenta \cite{piazza2016fflo} one finds $\rho(\omega) \propto \omega^{1/3}$. 
\item[-] \textit{specific heat capacity}: 
although the determination of thermodynamic quantities is a somewhat subtle issue (see Sec.\ \ref{concsec}), we expect that $C(T) \propto T^{\frac{4}{3} - \frac{2}{3}\theta}$. Here, $\theta = \theta(\delta v)$ is a hyperscaling violation exponent, which should fulfill $\theta(\delta v = 0) = 1$.
\end{itemize}
Finally, our RG analysis also identifies possible competing orders which may preempt the FFLO transition and lead to a ``competing order dome'' around the FFLO critical point. We find a charge density wave (CDW) peaked at $2k_F$ to be the most promising candidate. Since $2k_F$
is much larger than $Q_{\text{FFLO}}$,
an experiment sensitive to momentum (e.g., using x-ray scattering techniques) could serve to distinguish between the FFLO and CDW orders, although in practice difficulties may arise due to the required low temperatures and high magnetic fields \cite{Wosnitz}.

\section{Critical theory}
\label{crittheorysec}

\subsection{Critical theory in $2+1$ dimensions}

\begin{figure}
\centering
\includegraphics[width=.7\columnwidth]{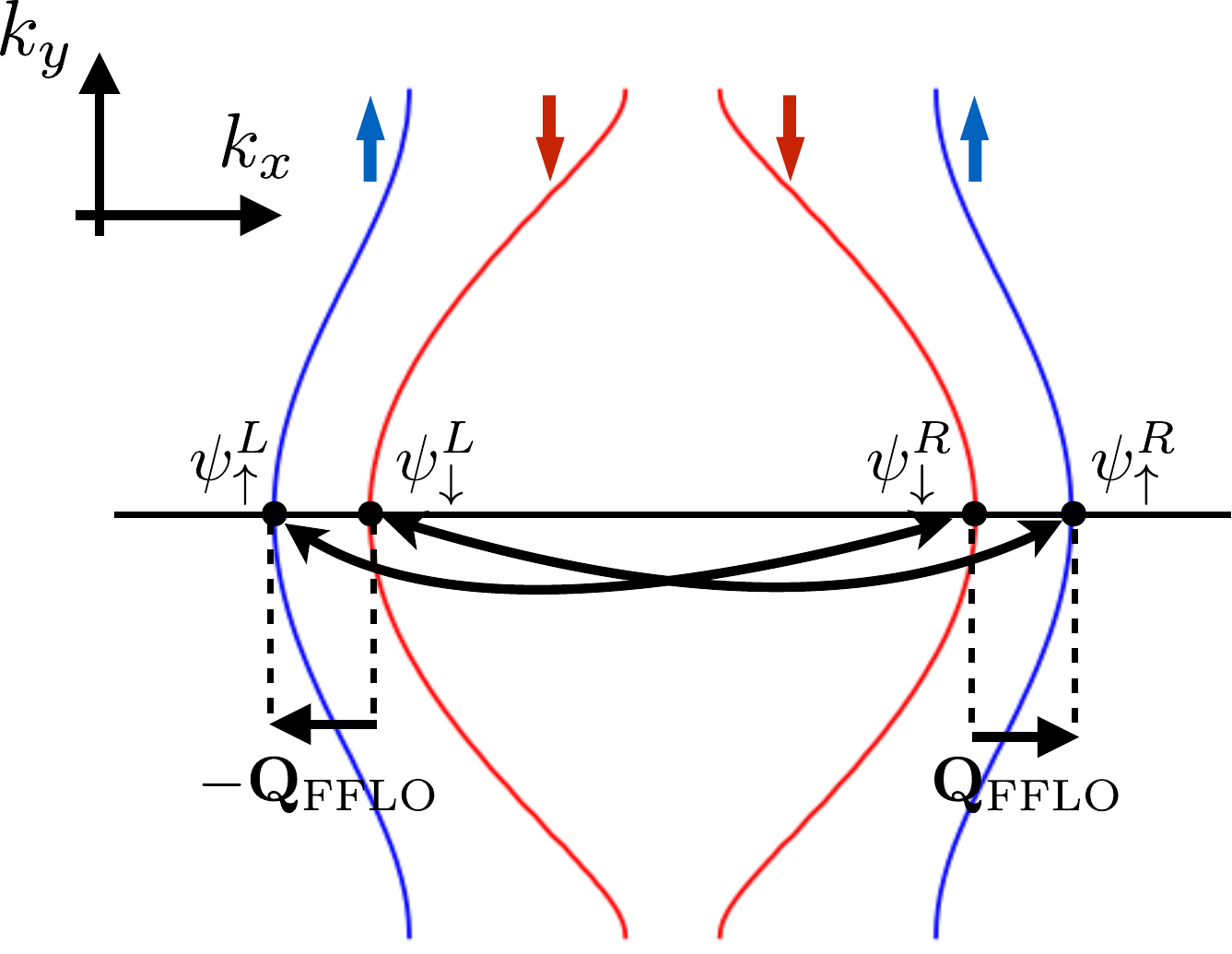}
\caption{(Color online) Typical Fermi surface of an anisotropic metal susceptible to FFLO pairing. Fluctuations of the pairing amplitude $\Delta$ strongly couple left branch fermions with right branch fermions with opposite spin at the hot spots. This picture was adapted from \cite{piazza2016fflo}.} 
\label{Fermisurface}
\end{figure}

When an anisotropic 2D metal at $T=0$ is subjected to a strong in-plane magnetic field $h$, and orbital effects can be neglected, the electron Fermi surface is spin-polarized. A typical sketch is shown in Fig.\ \ref{Fermisurface}. Let's now assume that the electrons interact with some generic short-range interaction 
\begin{align}
\label{Hint}
H_{\text{int}} = -g \int d^2 \r\  \psi_\uparrow^\dagger (\r )\psi_\downarrow^\dagger (\r ) \psi_\downarrow (\r) \psi_\uparrow (\r) \ ,  
\end{align}
resulting in Cooper pairing. To derive a low-energy effective action which makes this pairing explicit, one can perform an exact Hubbard-Stratonovich decoupling of the interaction term (\ref{Hint}) in the Cooper channel; thereby, one introduces bosonic fields $\Delta(\r), \bar{\Delta}(\r)$ with a free action $\int d\tau d^2 \r\ g |\Delta(\r)|^2$, which couple to the Fermions in Yukawa-like manner, $\propto g\Delta(\r) \overline\psi_\uparrow(\r)\overline\psi_\downarrow(\r)$. Due to the spin-polarization and the anisotropy of the Fermion dispersion, the bosonic fields (which correspond to the pairing amplitude) are peaked at momenta $\pm \textbf{Q}_{\text{FFLO}} \neq 0$, which is the very definition of the FFLO state. 
Due to the electron fluctuations, the bosonic mass term gets renormalized, $g \rightarrow m \equiv  g - \Pi(0,0;h)$, where $\Pi$ is the inverse pair propagator at vanishing energy-momentum, and we explicitly denoted its magnetic field dependence.\footnote{We perform the Hubbard-Stratonovich decoupling in such a way that $\Pi \propto g^2$, which is why our bare boson mass is $g$ instead of $1/g$.} As $h$ is increased above the Pauli upper critical field $h = h_c$, the renormalized mass changes from negative to positive values, and the system crosses from the FFLO phase to the normal metal phase along the $(T=0)$-line in the phase diagram of Fig.\ \ref{phasediag}. Accordingly, $m$ is proportional to the reduced magnetic field, $m \propto (h-h_c)/h_c$, in precise analogy to Ginzburg-Landau theory.  Further details on the procedure described above are presented in Appendix \ref{meanfieldapp}, illustrated by a mean field discussion of the phase transition for a specific microscopic model. 

By phase-space considerations, the low-energy fermions at the four hot spots with vanishing curvature in the $k_y$-direction shown in Fig.\ \ref{Fermisurface} are most strongly susceptible to pairing, with Cooper pair wave vectors $\pm \textbf{Q}_{\text{FFLO}}$. Following the above rationale, a zero temperature action which captures the phase transition between the FFLO and normal metal phases can be readily derived along the lines of Ref.\ \cite{piazza2016fflo} (see Eq.\ (4) therein):  
\begin{align}
\label{veryfirstaction}
\mathcal{S}  = &\int_{k^{2+1}}\! \sum_{\substack{i = L,R\\ \alpha = \uparrow, \downarrow}} \bar{\psi}^i_{\alpha}(k) \left(-ik_0 +  v^i_{\alpha} k_x +  k_y^2 \right)\psi_{\alpha}^i(k)\ 
\notag \\ & + \int_{k^{2+1}} (m + k_0^2 + k_x^2 + k_y^2)|\Delta(k)|^2 \notag \\ \notag &-g \int_{{k^{2+1}},{p^{2+1}}}\!\big[ \bar{\Delta}(k) \psi_{\downarrow}^L(p) \psi_{\uparrow}^R(k-p) \\ & + \bar{\Delta}(k)\psi_{\downarrow}^R(p) \psi_{\uparrow}^L(k-p) + \text{h.c.} \big] , 
\end{align}    
where $k_0  = \omega$, and 
 \begin{align}
& v_\alpha^i = \begin{cases} - v_\alpha \ , i = L \\ 
+v_\alpha \ , i = R \end{cases} \hspace{-1em},\  v_\alpha>0, \quad   \text{and} \quad \int_{k^{d+1}}\! \equiv \int\!\frac{d^{d+1} k}{(2\pi)^{d+1}}.
\end{align}
Here, the fermion fields $\psi^i_\alpha$ are expanded around the respective hot spots (see Fig.\ \ref{Fermisurface}), while the boson fields $\Delta$ are expanded around $\pm \textbf{Q}_{\text{FFLO}}$. For simplicity, we assume that the pairing is of Larkin-Ovchinnikov-type \cite{Larkin1965}, 
$\Delta(\r) \propto \cos(\textbf{Q}_{\text{FFLO}}\!\cdot \r)$, peaked around $\pm \textbf{Q}_{\text{FFLO}}$ with equal amplitude.

By the Hubbard-Stratonovich procedure sketched above, the bosons $\Delta$ originally just have a mass term $m \propto g$ and no dispersion. However, the kinetic terms and the renormalized mass will be automatically generated during the RG procedure, when high-energy degrees of freedom are integrated out (or, equivalently, arise from the leading analytical boson self-energy corrections involving fermions \cite{piazza2016fflo}).
Since an action which is appropriate for RG analysis should contain all analytical RG-relevant terms (non-analytical terms do not renormalize), we include these additional boson terms here from the start. Note that terms $(k_0^2 + k_x^2)|\Delta|^2$ are actually RG-irrelevant by tree-level power counting (see below), which is why we don't need curvature coefficients for them. 
Alternatively, one can just view the boson terms as expansion in powers and gradients of an FFLO pairing order parameter $\Delta$, as familiar from other non-Fermi liquid scenarios like Ising-nematic \cite{metlitski2010quantum} or SDW order \cite{Metlitski2010}. 

\subsection{Mean field analysis of superconducting phase} 
\label{meanfieldsec}

\begin{figure}
\centering
\includegraphics[width=\columnwidth]{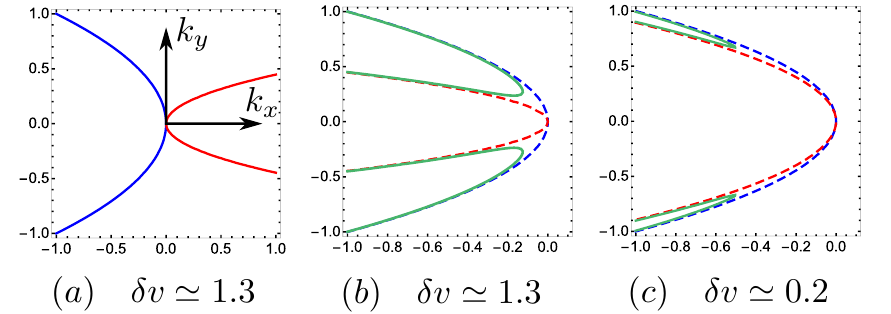}\caption{(Color online) Fermi surfaces in the FFLO phase at mean field level. (a) Fermi surface without pairing, i.e. $g\Delta_0 = 0$. (b) Fermi surface for electron/hole operators (see main text), at $\delta v \simeq 1.3$. Dashed lines: $g\Delta_0 = 0$. Full lines: $g\Delta_0 = 0.05$. (c) Same as (b), but $\delta v \simeq 0.2$.}
\label{meanfieldpic}
\end{figure}

As a first step, let us recall the mean-field level treatment of the action (\ref{veryfirstaction}) (compare, e.g., Refs.\ \cite{lutchyn2011spectroscopy,sheehy2007bec, kinnunen2017fulde}) in the superconducting phase, which amounts to the replacement $\Delta(k) \rightarrow \Delta_0 \cdot  \delta^{(3)}(k), |\Delta_0|>0$. For clarity, we focus on the fermionic branches $ \psi_{\uparrow}^R,\psi_{\downarrow}^L$, with dispersions   
\begin{align}
\xi_\uparrow^R(k) &= v_\uparrow k_x + k_y^2 \\ 
\xi_\downarrow^L(k) &= -v_\downarrow k_x + k_y^2. \notag
\end{align} 
A zoom-in on the respective Fermi surfaces  (compared to Fig.\ \ref{Fermisurface}, momenta are shifted towards a common origin) is shown in Fig.\ \ref{meanfieldpic}(a). The parameter which determines the Fermi surface shapes is the velocity detuning $\delta v$: 
\begin{align}
\label{veldef}
\delta v \equiv  2(v_\uparrow - v_\downarrow)/(v_\uparrow + v_\downarrow).
\end{align}
We now introduce Nambu-spinors in the standard fashion
\begin{align}
\Phi(k) = \left(\psi_{\uparrow}^R(k),\ \overline{\psi}_{\downarrow}^L(-k)\right)^T.
\end{align}
This means that we perform a particle-hole transformation for the spin-down electrons; the Fermi surface of the new fermionic degrees of freedom without pairing is shown in Fig.\ \ref{meanfieldpic}(b) (dashed lines). The mean field pairing Hamiltonian derived from Eq.\ (\ref{veryfirstaction}) is then readily diagonalized by Bogoliubov transformation, with rotated degrees of freedom 
\begin{align}
&\gamma_{+}(k) =  u_k \psi_{\uparrow}^R(k) - v_k \overline{\psi}_{\downarrow}^L(-k) \\
&\overline{\gamma}_{-}(k) =  v_k \psi_{\uparrow}^R(k) + u_k \overline{\psi}_{\downarrow}^L(-k).  
\notag 
\end{align}
where $u_k, v_k$ are some weights. The corresponding dispersions read: 
\begin{align}
E_{\pm} = & \pm \frac{1}{2}\bigg(\xi_\uparrow^R(k) -\xi_{\downarrow}^L(-k)  \\ &\pm \sqrt{(\xi_\uparrow^R(k) + \xi_{\downarrow}^L(-k))^2 + 4g^2|\Delta_0|^2}\bigg).  
\notag
\end{align}
Unlike in the BCS problem, gapless fermionic degrees of freedom remain; the ground state of the system is a condensate of Cooper pairs with a Fermi sea of $\gamma_{\pm}$ on top.
A plot of the corresponding $\gamma_+$ Fermi surface for $\delta v \simeq 1.3$ is shown in Fig.\  \ref{meanfieldpic}(b) (full green line); $\gamma_{-}$ fermions are gapped for $\delta v >0$.  

Microscopically, the parameter $\delta v$ grows monotonously for increasing magnetic fields. This parameter also controls the effectiveness of pairing. Indeed, for $\delta v \rightarrow 0$ the full Fermi surface gaps out; the problem becomes BCS-like. This trend is demonstrated in Fig.\ \ref{meanfieldpic}(c), which shows the same quantities as Fig.\ \ref{meanfieldpic}(b), but for a significantly smaller value $\delta v \simeq 0.2$.

As already seen in Fig.\ \ref{meanfieldpic}, the limit $\delta v \rightarrow 0$ is rather peculiar. Still, taking this limit will be required later on to gain analytical control over the problem. The implications of this procedure will be discussed in more detail below [(see Sec.\ \ref{diagsec} and Appendix \ref{suplogApp})].

\subsection{Critical theory in $d+1$ dimensions}
\label{d+1critsec}

Let us now focus on the phase transition from the FFLO to the normal metal phase, which can be driven by tuning the boson mass $m$ in Eq. (\ref{veryfirstaction}) from negative to positive values. Going beyond a Landau-Ginzburg type analysis of the phase transition (as found, e.g.,  in Refs.\ {\cite{buzdin1997generalized,radzihovsky2009quantum}), we will treat both bosons and fermions as dynamical degrees of freedom, and look for the critical RG fixed point of the action (\ref{veryfirstaction}) in the IR. 
However, this fixed point is located at strong coupling; to access it perturbatively, we must introduce a small parameter $\epsilon$ into the action which suppresses quantum fluctuations. A convenient way of doing so is to increase the space dimension $d$, thereby successively tuning the Yukawa-interaction between bosons and fermions marginal as $d$ approaches the critical dimension $d_c$. For $d=d_c$, the interacting critical fixed point then collapses with the non-interacting Gaussian one, and we can therefore derive RG flow equations perturbatively in $\epsilon = d_c - d$.  

In the presence of a Fermi surface, one may increase the number of dimensions tangential or perpendicular to it \cite{mandal2015ultraviolet, *mandal2016uv,lee2017recent}. 
Some aspects of the scheme with increased tangential dimensions (or fixed co-dimension), where $d_c = 3$, are outlined in Appendix \ref{codimApp}; in short, this extension is problematic because it leads to a breakdown of the hot spot theory in the parameter regime where the computations are analytically tractable. 
Let us therefore follow \cite{senthil2009fermi, dalidovich2013perturbative} and increase the perpendicular dimensions. I.e., the Fermi surface is always one-dimensional, and the fermionic density of states is succesively reduced. This amounts to an expansion around $d_c = 5/2$.

To implement this dimensional extension in practice, we employ the formalism and techniques introduced in Ref.\ \cite{dalidovich2013perturbative}, where renormalization group equations are computed within the dimensional regularization (called dimReg henceforth) and minimal subtraction schemes (see Refs.\ \cite{kleinert2001critical,vasil2004field} for an introduction). We will work at $T=0$; thermal fluctuations on a different, isotropic model for the FFLO transition were recently studied in Ref. \cite{thermal_fflo_rg_Jakub} with functional RG methods.

 For shorter notation, we define fermionic  ``spinors" $\Psi$: 
\begin{align}
\label{spinorrep}
\Psi_\alpha(k) = \begin{pmatrix}
\psi_{\alpha}^R(k)\\ \bar{\psi}_{\alpha}^L(-k) 
\end{pmatrix} , 
\ \bar{\Psi}_\alpha(k) = \begin{pmatrix}
\bar{\psi}_{\alpha}^R(k),\!& \psi_{\alpha}^L(-k) 
\end{pmatrix} \cdot \sigma_y,  
\end{align}
where $\sigma_y$ is a Pauli matrix. The  kinetic term for the fermions can then be generalized to $d+1$ dimensions as
\begin{align}
\label{fermionkinetic}
\sum_{ \alpha = \downarrow,\uparrow}\int_{k^{d+1}}  \bar{\Psi}_{\alpha}(k) \left(-i\boldsymbol{\Gamma}\cdot \boldsymbol{K} + i \sigma_x \delta_\alpha(k)  \right)\Psi_{\alpha}(k). 
\end{align}
Here, $\boldsymbol{K} = (k_0, k_1, ... , k_{d-2})$, and the momenta $k_x, k_y$ are relabeled as $k_x \! \rightarrow \! k_{d-1}, k_y \!  \rightarrow \! k_d$. $\delta_\alpha$ is the right branch fermion dispersion, $\delta_\alpha = v_\alpha k_{d-1} + k_d^2$. 
$\boldsymbol{\Gamma} =  (\gamma_0, \gamma_1, ..., \gamma_{d-2})$ is a vector of two-dimensional Gamma-matrices which fulfill the Clifford algebra, $\{\gamma_\alpha,\gamma_\beta\} = 2\delta_{\alpha\beta}$. In the integer cases of interest:  
\begin{align}
& d=2: \quad  \boldsymbol{K} = k_0,\quad  \boldsymbol{\Gamma} = \sigma_y \\
& d=3: \quad  \boldsymbol{K} = (k_0,k_1), \quad \boldsymbol{\Gamma} = (\sigma_y, \sigma_z). 
\end{align}
To uniquely specify the Gamma-matrix structure, in general dimensions we choose the continuation
\begin{align}
\label{Gammavec}
\boldsymbol{\Gamma} = (\sigma_y, \vec{\Gamma}) = (\sigma_y, \sigma_z, \hdots \sigma_z), 
\end{align}
where the ``vector" $\vec{\Gamma}$ has $(d-2)$ entries. 

The introduction of generalized 
Gamma-matrices is a standard tool in dimReg of fermionic theories, see e.g.\ \cite{Peskin1995}. In the condensed matter context, an alternative  point of view is the following: we add \textit{one} extra dimension perpendicular to the Fermi surface, extending the action with terms of triplet-pairing form \cite{dalidovich2013perturbative}
\begin{align}
\label{anomalousGF}
\bar{\Psi}_{\alpha}(k) (-ik_1 \sigma_z) \Psi_\alpha(k) = 
k_1\bigg(\!\sum_{\alpha = \downarrow,\uparrow} \bar{\psi}_\alpha^R(k)\bar{\psi}_{\alpha}^L(-k) + \text{h.c.} \bigg). 
\end{align}
These terms gap out the Fermi surface except for the one-dimensional branches of Fig.\ \ref{Fermisurface}. In all computations, we then continuously tune the ``weight" of this extra dimension, by using a radial integral measure $\int \!dk_1k_1^{d-3} /(2\pi)^{(d-2)} $. It should be noted that by introducing these extra terms we have broken the spin-rotation symmetry in the $xy$-plane  of the original action (\ref{veryfirstaction}). 

The kinetic term for the bosons in the $(d+1)$-dimensional action generalizes to
\begin{align}
\label{bosonkinetic}
\int_{k^{d+1}} (|\boldsymbol{K}|^2 + k_{d-1}^2 + k_d^2 + m)|\Delta(k)|^2.
\end{align}

The terms in the noninteracting parts of the action (\ref{fermionkinetic}),(\ref{bosonkinetic}) are invariant under the scaling transformations 

\begin{align}
\label{scaletrafo}
&\boldsymbol{K} = \frac{\boldsymbol{K}^\prime}{b} \ , \quad k_{d-1} = \frac{k^\prime_{d-1}}{b},\ \quad k_d = \frac{k_d^\prime}{\sqrt{b}} \\  &\Psi(k) = b^{\tfrac{d}{2} + \tfrac{3}{4}}\Psi^\prime(k^\prime) \ , \quad \Delta(k) = b^{\tfrac{d}{2} + \tfrac{3}{4}} \Delta^\prime(k^\prime). \notag
\end{align}
At tree level, the terms $(|\K|^2 + k_{d-1}^2 )|\Delta|^2$ are irrelevant, and will stay so in $\epsilon$-expansion as long as $\epsilon$ is small. Let's therefore erase these terms from the action. Furthermore, as we are mostly interested in the quantum critical point, we will set the renormalized mass $m=0$ in the following. The IR divergences resulting from these two steps can be regularized by using dressed boson propagators in all computations \cite{dalidovich2013perturbative}. 

Inserting the spinor definitions (\ref{spinorrep}), the interaction term is easily rewritten in higher dimensions. In total, the critical action in $d+1$ dimensions then reads 
\begin{align}
\label{critactionhigherdim} \notag
\mathcal{S} =& \int_{k^{d+1}}  \bar{\Psi}_{\alpha}(k) \left(-i\boldsymbol{\Gamma}\cdot \boldsymbol{K} + i \sigma_x(v_\alpha k_{d-1} +  k_d^2) \right)\Psi_{\alpha}(k) \\  &+ \int_{k^{d+1}}  k_d^2 \cdot  |\Delta(k)|^2 \notag \\ &- g \mu^{\epsilon/2}\int_{k^{d+1}, p^{d+1}}  \big[\bar{\Delta}(k) \sigma_y^{\alpha  \alpha^\prime} \bar{\Psi}_\alpha(-p) M_1 \Psi_{\alpha^\prime}(k-p) \notag \\ &+ \Delta(k) \sigma_y^{\alpha \alpha^\prime} \bar{\Psi}_{\alpha} (k-p) M_2 \Psi_{\alpha^\prime}(-p)\big], 
\end{align}
where we introduced matrices acting in spinor space
\begin{align}
\quad   M_1 = \begin{pmatrix} 1 & 0 \\ 0 & 0 \end{pmatrix} , \quad M_2 = \begin{pmatrix} 0 & 0 \\ 0 & 1 \end{pmatrix}, \label{Mmatrices} 
\end{align}
and employed a summation convention for spin indices. Note that the pairing terms of the original action (\ref{veryfirstaction}) have the form of a standard density term in the spinor language. 
We have made the tree-level scaling dimension of the interaction explicit by replacing $g\rightarrow g\mu^{\epsilon/2}$, where $\mu$ is an arbitrary mass scale, and 
\begin{align}
\epsilon = 5/2 -d . 
\end{align}  
In the standard logic of $\epsilon$-expansion, we will work in the limit $\epsilon \rightarrow 0$, where the interaction term becomes marginal, and determine the critical exponents at the interacting fixed point to order $\epsilon$. 
Extrapolating to the physically relevant value $\epsilon = 1/2$, we can then make a controlled qualitative estimate of critical exponents and the universality class of the problem.

\section{One-loop diagrams} 
\label{diagsec}

To compute the flow equations in dimReg, one needs to evaluate the possible one-loop corrections to the action (\ref{critactionhigherdim}), whose diagrammatic representations are shown in Fig.\ \ref{oneloopdiags}. Note that tadpole contributions to the fermion self-energy are disregarded since they can renormalize the chemical potential only. Higher loop-digrams are multiplied with a higher power of the coupling $g$. Below, we will show that $g \propto \epsilon^{3/4}$ at the critical point, thus higher-loop diagrams are suppressed for $\epsilon \rightarrow 0$. In this work, we will disregard them alltogether.

\begin{figure}
\centering
\includegraphics[width=.6\columnwidth]{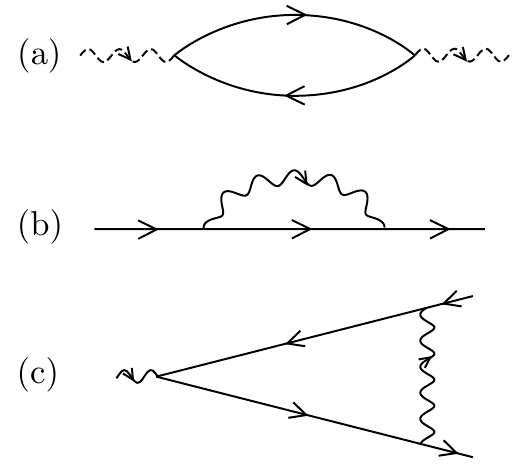}
\caption{One-loop diagrams. Dashed wavy lines (Fig.\ (a)) indicate bare boson propagators, while straight lines indicate electron propagators. Full wavy  lines (Figs.\ (b), (c)) represent bosons dressed with the self-energy of Fig. (a). External lines are amputated. }
\label{oneloopdiags}
\end{figure}

To evaluate these diagrams analytically, we need to make one important approximation: We consider the limit of \textit{vanishing velocity detuning}, $\delta v \rightarrow 0$ [c.f.\ Eq.\ (\ref{veldef})]. 

In a realistic experimental setup, $\delta v = \mathcal{O}(0.1)$ \cite{piazza2016fflo} is indeed small. However, the limit $\delta v \rightarrow 0$, while being computationally convenient, is somewhat singular, as already indicated in Sec.\ \ref{meanfieldsec}.  This can be seen pictorially in Fig.\ \ref{singularlimitpatches2}: for nonvanishing velocity detuning [Fig.\ \ref{singularlimitpatches2}(a)], two Fermi surface branches interacting with each other have different curvatures. Thus, only electrons with momenta close to the hot spot at $\k=0$ (the branches are shifted towards a common origin) scatter strongly with FFLO fluctuations. For any electron close to the Fermi surface with large momentum $\k$ away from the hot spot  [red dot in Fig. \ref{singularlimitpatches2}(a)], the corresponding electron with momentum $-\k$ (indicated by a dashed line and a blue dot), which would be most susceptible to FFLO pairing, has momentum far from the Fermi surface, and thus pairing is suppressed. 

On the other hand, if the two spin-velocities are equal [Fig.\ \ref{singularlimitpatches2}(b),(c)], an arbitrary electron on the Fermi surface with momentum $\k_1$ can scatter against its counterpart with momentum $-\k_1$, as also demonstrated in Sec.\ \ref{meanfieldsec}. However, the FFLO fluctuations can only scatter these electrons efficiently into a pair of electrons with momenta $\pm \k_2$, s.t. $\k_2 \simeq \k_1$. The tangent vector to the Fermi surface of the initial pair, $\q_1$, must almost coincide with the final tangent vector, $\q_2$, as shown in Fig.\ \ref{singularlimitpatches2}(b). If $\q_2 \not\simeq \q_1$, as shown in Fig.\ \ref{singularlimitpatches2}(c), the scattering process is energetically suppressed. The fact that scattering processes are only local in momentum space prevents the explicit appearance of UV scales and thereby justifies application of the hot spot theory. Note that this argument remains true only as long the Fermi surface is strictly 1D; for higher dimensional Fermi surfaces, which arise in the RG scheme with fixed co-dimension, the limit $\delta v \rightarrow 0$ is even more singular and results in UV-IR mixing \cite{mandal2015ultraviolet}, eventually leading to a break-down of the hot spot expansion; see Appendix \ref{codimApp} for further details.

\begin{figure}
\centering
\includegraphics[width=\columnwidth]{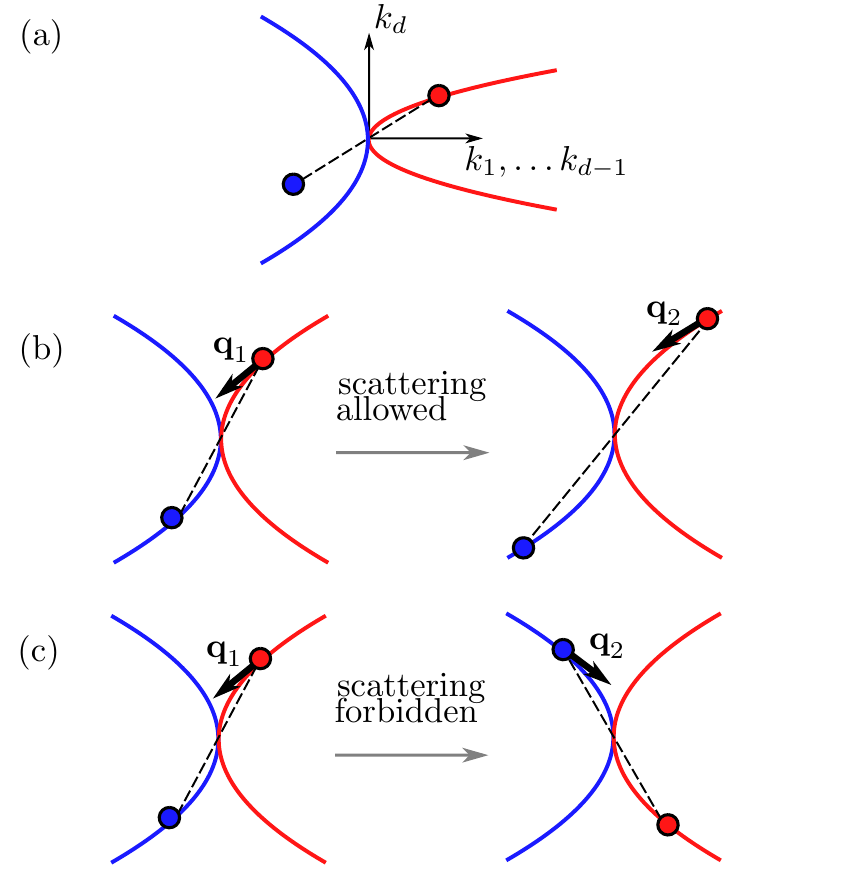}
\caption{(Color online) Zoom-in of Fig.\ \ref{Fermisurface}, showing the two Fermi surface branches of fermions coupled by pairing fluctuations at $+ \textbf{Q}_{\text{FFLO}}$, shifted towards a common origin in momentum space.
(a) Case of nonzero velocity detuning $\delta v \neq 0$, where only fermions with momenta close to $k = 0$ are strongly entangled. 
(b) Case of vanishing velocity detuning $\delta v \rightarrow 0$, where the electrons with momenta $\pm \k_1$ and tangent vector $\q_1$ can be scattered to close-by momenta $\pm \k_2$ with similar tangent vector $\q_2$. (c) same initial configuration as Fig. (b), but with different final momenta $\pm \k_2$; now, the final tangent vector $\q_2$ differs strongly from the initial one, and the phase space for the scattering is negligible. } 
\label{singularlimitpatches2}
\end{figure}

Despite its smallness, in a fully fledged RG analysis of the problem $\delta v$ should be treated as a running coupling. We will leave this involved task for future (numerical) work, and focus on $\delta v \rightarrow 0$ from now on, which should be qualitatively correct as long as $\delta v$ does not exhibit a runaway flow in the full RG procedure.  

Let's now evaluate the boson self-energy $\Pi$ of Fig.\ \ref{oneloopdiags}(a). This diagram dresses the bare boson Green's function 
\begin{align}
D_0(k) \equiv  \braket{\Delta(k) \bar{\Delta}(k)}_0 = 1/k_d^2, 
\end{align}
where the subscript $0$ indicates that averages are taken with respect to the noninteracting action, 
and reads:
\begin{align}
\label{Bose selfenergy}
\Pi(k)  = -g^2\mu^{\epsilon} \int_{p^{d+1}}\sum_{\alpha \neq \alpha^\prime} \text{Tr}\left[G_\alpha (-p) M_1 G_{\alpha^\prime}(k-p) M_2\right].
\end{align} 
Here, the electron Green's function is defined by 
\begin{align}
\label{DLGF}
G_{\alpha} (k) \equiv \braket{\Psi_\alpha(k) \bar{\Psi}_\alpha(k)}_0 = -i \frac{-\boldsymbol{\Gamma} \cdot \boldsymbol{K} + \sigma_x \delta_k}{\boldsymbol{K}^2 + (\delta_k)^2}, 
\end{align}
where $\delta_k = k_{d-1} + k_d^2$, i.e., we have scaled out the equal velocities. Evaluation of (\ref{Bose selfenergy}) is done in Appendix \ref{BoseSelfApp}, and yields
\begin{align}
\label{PikfinalDL}
\Pi(k) = \chi_d \frac{g^2\mu^{\epsilon}}{|k_d|} (d \cdot k_0^2+|\vec{k}|^2) \cdot (k_0^2 + |\vec{k}|^2)^{\tfrac{d-3}{2}}  \ ,
\end{align}
with 
\begin{align}
\chi_d &= \frac{\Gamma((1 -d)/2)}{2^{d+2}\pi^{(d+1)/2} }\cdot \frac{\Gamma(d/2)^2}{\Gamma(d)}, \\
\chi_{5/2} &\simeq -0.0178. \notag 
\end{align}
In Eq.\ (\ref{PikfinalDL}), $\vec{k}$ are the extra dimensions inserted in the dimReg scheme, i.e. $\boldsymbol{K} = (k_0, \vec{k})$. The fact that we have an anisotropy in $\boldsymbol{K}$-space is a peculiarity of the original pairing vertex, leading to a matrix structure in spinor space with matrices $M_1, M_2$ [see Eqs.\ (\ref{critactionhigherdim}),(\ref{Mmatrices})], which are not Gamma-matrices. This anisotropy can be easiest understood taking the fermion self-energy as an example, see below. For $d=2$, there are no extra dimensions, and Eq.\ (\ref{PikfinalDL}) simplifies to the 2D result found in Ref.\ \cite{piazza2016fflo}. 

Two further comments on the result (\ref{PikfinalDL}) are in order. First: To arrive at (\ref{PikfinalDL}), we had to make a trivial regularization by subtracting $\Pi(0,k_d)$ (in any dimension). The residual momentum dependence of this subtraction is an artefact of the $\delta v\rightarrow 0$ limit: for $\delta v \neq 0$, at least in the physical case $d=2$, one obtains a finite result for the self-energy by subtracting $\Pi(\delta v \neq 0, k = 0)$. If we could take $\delta v \rightarrow 0$ in the last step of the computation, i.e.\ before dropping momentum cutoffs, this trivial mass renormalization (which is perfectly legitimate as we focus on the critical point where the boson is massless) would always suffice. However, in practice we have to take the limit $\delta v\rightarrow 0$ first, and subtract $\Pi(0,k_d)$ (which amounts to a ``superconducting logarithm") in effect. A more detailed justification of this step is presented in Appendix \ref{suplogApp}. 
Second: Although at first glance of the Fermi surface of Fig.\ \ref{Fermisurface} one could expect $\Pi$ to have a SDW-type behaviour $\Pi(k) \sim |\boldsymbol{K}|$ \cite{Metlitski2010},  our result  (\ref{PikfinalDL}) is a standard Landau-damping term familiar from the Ising-nematic case \cite{dalidovich2013perturbative}, apart from the anisotropy discussed above. This is again a consequence of the pairing structure of the original vertex.

As in the Ising-nematic case, the boson self-energy is UV finite as $d\rightarrow 5/2$. Still, this contribution is crucial, as the further loop corrections of Fig. \ref{oneloopdiags} (b),(c) are only IR finite if the boson lines are taken to be dressed, which we will do in the following, compare Ref. \cite{dalidovich2013perturbative}. 

Let's now evaluate the fermion self-energy of Fig.\ \ref{oneloopdiags} (b). For a fermion of spin $\kappa$, there are two contributions
\begin{align}
\label{Sigma_1}
&\Sigma_1^\kappa(k) = g^2 \mu^{\epsilon} \int_{p^{d+1}} D(p) M_1 G_\beta (k+p) M_2 \sigma_y^{\kappa \beta} \sigma_y^{\beta \kappa} \ , \\ &
\label{Sigma_2}
\Sigma_2^\kappa(k) = g^2 \mu^{\epsilon} \int_{p^{d+1}} \hspace{-.7em} D(-p) M_2 G_\beta (k+p) M_1 \sigma_y^{\kappa \beta} \sigma_y^{\beta \kappa} \ , 
\end{align}
representing the two ways to draw the arrow on the boson line. Evaluating these integrals in leading order in $\epsilon$ (see Appendix \ref{FermiSelfApp}), we obtain 
\begin{align} 
\label{fermiselfmaintext}
\Sigma(k) &= \Sigma_1^\kappa(k) + \Sigma_2^\kappa(k) \\& =  
  \frac{u_g g^{4/3}}{\epsilon}   \cdot \sigma_y \cdot (-ik_0) + \text{finite terms},  \notag \\ 
u_g &\simeq -0.0813. \notag   
\end{align}
Thus, we find that the fermion self-energy only depends on the frequency, and not on the extra momenta $\vec{k}$ as for the Ising-nematic \cite{dalidovich2013perturbative}. This is easily understood as follows: as discussed before, see Eq.\ (\ref{anomalousGF}), insertion of extra dimensions $\vec{k}$ gives rise to triplet pairing terms already at the noninteracting level, or, in other words, to anomalous terms in the bare fermion Green's function $\propto \vec{k}$, when expressed in terms of the original fermion fields $\psi$ (see, e.g., Ref.\ \cite{G.D.Mah2000}). Therefore, to obtain a contribution to $\Sigma(k) \propto \vec{k}$, there must be an anomalous contribution to the self-energy. However, this is not possible at one-loop. This is seen pictorially in Fig.\ \ref{nokfig}(a), which shows an impossible diagram (since four fermions are annihilated at the vertices) in terms of original fermion fields. Note that at higher loop level such contributions can arise, see Fig.\ \ref{nokfig}(b). 

\begin{figure}
\centering
\includegraphics[width=\columnwidth]{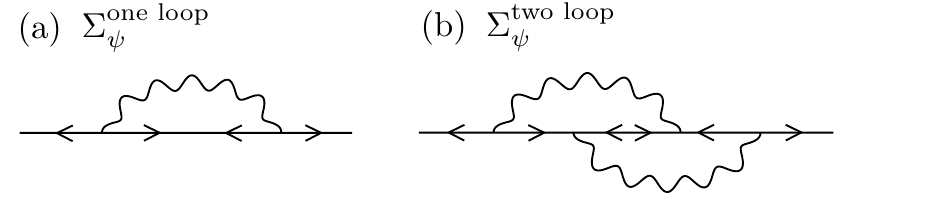}
\caption{Anomalous contributions to the fermion self-energy (in terms of the original fermion fields $\psi$). At one-loop, such contributions are impossible [Fig.\ (a)], but can arise at two loop [Fig.\ (b)]. } 
\label{nokfig}
\end{figure}

Last, we need to compute the vertex correction of Fig.\ \ref{oneloopdiags}(c). In $d=2$, this diagram is trivially absent, but not in $d>2$ (due to the anomalous terms). However, we still find that there is no $\epsilon$-divergent vertex correction; further details are relocated to Sec.\ \ref{compordsec}, where we discuss general vertex corrections that reflect possible competing orders.

\section{Renormalization}

\label{rensec}

\subsection{Flow equation}

To obtain a UV finite renormalized action, we have to add the fermion self-energy as a counter-term, employing the minimal subtraction scheme where the counterterm depends on $g$ only: 
\begin{align}
\label{SCT}\notag 
S_{\text{CT}} = &\sum_{\alpha = \downarrow, \uparrow}\int_{k^{d+1}} \frac{Z_{1,1}(g)}{\epsilon} \bar{\Psi}_{\alpha}(k)(-i \sigma_y \cdot k_0)\Psi_{\alpha}(k), \\  & Z_{1,1}(g) = u_g \cdot g^{4/3}.
\end{align}
Then, the renormalized action is obtained as $\mathcal{S}_{\text{ren}} = \mathcal{S} + \mathcal{S}_{\text{CT}}$. We define a renormalization constant $Z_1 = 1+ Z_{1,1}/\epsilon$ and introduce unrenormalized (bare) fields and couplings as 
\begin{align}
\label{momscalingz}
& k^b_{0} = k_0\cdot Z_1, \quad  \vec{k}^b = \vec{k},  \quad k^b_{d-1} = k_{d-1}, \quad k^b_{d} = k_d,  \\ \notag &\Psi_b(k_b) = Z_1^{-1/2} \Psi(k), \quad \Delta_b(k_b) = Z_1^{-1/2} \Delta(k), \\  & g^b = Z_1^{-1/2} \mu^{\epsilon/2} g.
\notag 
\end{align}
These relations bring the renormalized action back in the form of the initial bare action (\ref{critactionhigherdim}) except for the dimensionful coupling $g^b$:
\begin{align}  
\label{bareaction} \notag 
&\mathcal{S}_{\text{ren}} = \\ \notag & \int_{(k^b)^{d+1}} \! \bar{\Psi}^b_{\alpha}(k^b) \left(-i\boldsymbol{\Gamma}\cdot \boldsymbol{K}^b + i \sigma_x( k^b_{d-1} +  (k^b_d)^2) \right)\Psi^b_{\alpha}(k^b) \\  &+ \int_{(k^b)^{d+1}}  (k_d^b)^2 \cdot  |\Delta^b(k^b)|^2 \notag \\ &- g^b \hspace{-2.5em}\int\displaylimits_{(k^b)^{d+1}, (p^b)^{d+1}} \hspace{-2em} \big[\bar{\Delta}^b(k^b) \sigma_y^{\alpha  \alpha^\prime} \bar{\Psi}^b_\alpha(-p^b) M_1 \Psi_{\alpha^\prime}^b(k^b-p^b) \notag \\ &+ \Delta^b(k^b) \sigma_y^{\alpha \alpha^\prime} \bar{\Psi}^b_{\alpha} (k^b-p^b) M_2 \Psi^b_{\alpha^\prime}(-p^b)\big]. 
\end{align}
Let's determine the flow of the renormalized coupling $g$ at a fixed UV value of the bare coupling $g^b$ as the mass scale $\mu$ is decreased. It is described by the beta function
\begin{align}
\beta = \frac{dg}{d\ln(\mu)}, 
\end{align}
which fulfills the equation
\begin{align}
\label{betaeq}
\beta\left(\frac{g}{2}Z_1^\prime - Z_1\right) - \frac{\epsilon}{2} g Z_1 = 0.
\end{align} 
We may solve it making the standard ansatz $\beta = \beta_0 + \epsilon \beta_1$, where $\beta_{0,1}$ depend on $g$ only. Comparing coefficients of the parts regular in $\epsilon$ of Eq.\ (\ref{betaeq}) yields\footnote{Note that the solution (\ref{betasolution}) violates Eq.\ (\ref{betaeq}) at order $g^{11/3}/\epsilon$. This is a standard artefact of approximating the renormalization constant $Z_1 \simeq 1 + Z_{1,1}$ at one-loop level, and should be succesively improved by higher loop contributions.} 
\begin{align}
\label{betasolution}
\beta = -\frac{u_g}{3} g^{7/3} - \frac{\epsilon}{2} g. 
\end{align}
The beta function has a fixed point at
\begin{align}
g^\star = \left(\frac{3
\epsilon}{-2u_g}\right)^{3/4}, \quad  
u_g \simeq -0.0813.
\end{align}
Writing $\mu = \mu_0 e^{-\ell}$, the RG eigenvalue of $g$ at $g=g^\star$ in the IR ($\ell \rightarrow \infty$) is $-\tfrac{2}{3}\epsilon$, i.e, the fixed point is stable (respectively, critical, as we have dropped the RG relevant mass term from the action). This indicates a second order phase transition between the FFLO and normal metal phases. A continuous transition was also found in the mean-field study of our precursor work \cite{piazza2016fflo}, and other 2D studies \cite{burkhardt1994fulde,mora2004nature}.

\subsection{Critical properties}
\label{critpropsec}

Let's discuss critical properties of this new fixed point, which are intimately linked with experimental observables. First, we define the dynamical critical exponent $z$: 
\begin{align}
z = 1 - \frac{d\ln(1/Z_1)}{d\ln(\mu)} = 1 + \frac{1}{Z_1} Z_1^\prime \beta.
\end{align}
At the fixed point we find 
\begin{align}
\label{zstar}
z^\star = 1+ \epsilon.
\end{align}
From the renormalization of fields in Eq.\ (\ref{momscalingz}), the anomalous dimensions of bosons and fermions read 
\begin{align}
\label{etas} 
\eta_\Psi = \eta_\Delta = \frac{1}{2} \frac{d\ln (1/Z_1)}{d\ln (\mu)} = \frac{1-z}{2} = -\frac{\epsilon}{2}\bigg|_{z = z^\star}.
\end{align}
$z$ and $\eta$ feed into the scaling behaviour of correlation functions, which can be determined in the standard way, defining renormalized Green's functions by
\begin{align} 
\label{Gren}
\notag 
& \langle \Psi(k_1) \hdots \Psi(k_m) \bar{\Psi}(k_{m+1}) \hdots \bar{\Psi}(k_{2m}) \Delta(k_{2m+1})\hdots \\& \notag \Delta(k_{2m+n})\bar{\Delta}(k_{2m+n+1})\hdots \bar{\Delta} (k_{2m+2n})\rangle = \\& G^{(m,m,n,n)}(\{k_i\};g,\mu)  \notag \times \\ &\delta^{d+1}\big(\sum_{i=1}^m \!k_i + \!\sum_{i = 2m+1}^{2m+n} \! k_i- \!\sum_{j=m+1}^{2m}\! k_j - \! \sum_{j=2m+n+1}^{2m+2n}\! k_j\big), 
\end{align}
with spin and spacetime indices suppressed. These correlators are related to the bare ones derived from the bare action (\ref{bareaction}) by multiplicative renormalization, and fulfill the scaling equation 
\begin{align}
\label{GFscalingeq}
\nonumber
\bigg\{ &\sum_{i=1}^{2m+2n} z\cdot k_{i,0}\frac{\partial}{\partial k_{i,0}} + \vec{k}_i\nabla_{\vec{k}_i} + k_{i,d-1}\frac{\partial}{k_{i,d-1}} + \frac{k_{i,d}}{2} \frac{\partial}{\partial k_{i,d}} \notag \\ &- \beta \frac{\partial}{\partial g}  - 2m \cdot (\eta_\Psi - \tfrac{4-\epsilon}{2})- 2n \cdot (\eta_\Delta - \tfrac{4-\epsilon}{2})  \notag \\ &+ (\epsilon - z - 2)\bigg\} G^{(m,m,n,n)}(\{k_i\};g,\mu) = 0. 
\end{align}
At the fixed point where $\beta =0$, and the RG exponents are given in Eqs.\ (\ref{zstar}) and (\ref{etas}),   
Eq.\ (\ref{GFscalingeq}) implies a scaling form of the fermion two-point  function
\begin{align}
\label{twopointscaling}
G(\k) \propto \frac{1}{\delta_k}\cdot f\bigg(\frac{k_0^{1/(1+\epsilon)}}{\delta_k} , \frac{|\vec{k}|}{\delta_k} \bigg), \quad \delta_k = k_{d-1} + k_d^2,  
\end{align} 
where $f$ is a universal scaling function. 
In particular, in $d=2$ ($\epsilon = 1/2$), this scaling form is consistent with the fermion self-energy $\propto k_0^{2/3}$ obtained in Ref.\  \cite{piazza2016fflo}. We therefore find, for $\epsilon > 0$, non-Fermi liquid behaviour where the quasiparticle nature of fermions is destroyed by strong order parameter fluctuations; exactly at $\epsilon = 0$, the system is a marginal Fermi liquid. 
For bosons, one finds the same scaling form as in Eq.\ (\ref{twopointscaling})  with $\delta_k$ replaced by $k_d^2$: 
\begin{align}
\label{Dtwopointscaling}
D(\k) \propto \frac{1}{k_d^2}\cdot f\bigg(\frac{k_0^{1/(1+\epsilon)}}{k_d^2} , \frac{|\vec{k}|}{k_d^2} \bigg).
\end{align}
Apart from the critical correlations 
(\ref{twopointscaling}), also the scaling behaviour on the normal metal side is of interest, characterized by the correlation length exponent $\nu$. To find it, we need to include a mass perturbation $m|\Delta|^2$ in the action, and $\nu$ is given by the inverse RG eigenvalue of $m$. Then, we need to compute the boson self-energy $\Pi(0)$ -- the mass will aquire an anomalous dimension if $\Pi(0)$ shows a (logarithmic) $1/\epsilon$ divergence. In our evaluation of $\Pi$ in App.\ \ref{BoseSelfApp}, such a logarithmic divergence does not arise, at least at one-loop in the analytically controlled limit $\delta v \rightarrow 0$.  By power counting, we can thus conclude 
\begin{align}
\label{nuestimate}
\nu = 1 + \mathcal{O}(\epsilon^2).
\end{align}
What is more, our theory is similar to the nematic case, where the boson self-energy does not diverge up to 3 loop,   \cite{dalidovich2013perturbative}. So, we can expect that the estimate (\ref{nuestimate}) holds to higher loop level as well.

\section{Physical Observables}

Eqs.\ (\ref{twopointscaling}), (\ref{Dtwopointscaling}), obtained in a controlled perturbative procedure, are the major result of this work. Eq.\ (\ref{Dtwopointscaling}) tells us the scaling form of the pair susceptibility $D$. For ordinary BCS \cite{ferrell1969fluctuations, scalapino1970pair,PhysRevLett.25.743} as well as unconventional high-$T_c$ \cite{PhysRevB.84.144527} superconductors, the imaginary part of this quantity is proportional to the Josephson current in a SIN junction setup for a small applied bias voltage; it remains to be seen if this idea can be carried over to FFLO superconductors. Furthermore, by integration over $D^2$ (see App.\ \ref{suscapp}), one can obtain the fluctuation contribution to the spin susceptibility $\chi_\Delta$ in the normal state. For $d=2$, we find a weakly divergent behaviour as function of the reduced magnetic field, $\chi_\Delta \propto \ln ((h-h_c)/h_c)$. This is in agreement with the RPA result of Ref.\ \cite{samokhin2006quantum}.  

The correlator $G$ in Eq.\ (\ref{twopointscaling}) describes the fate of electronic excitations. In $d=2$, they decay in non-Fermi liquid manner, with a large rate $\Gamma(k_0) \propto k_0^{2/3}$. The hot-spot density of states $\rho(k_0)$ of these excitations can be found by integrating the electronic spectral function over momenta \cite{piazza2016fflo}, $\rho(k_0)\propto k_0^{1/3}$. In addition, a constant contribution to $\rho(k_0)$ from the cold, Fermi-liquid-like parts of the Fermi surface will arise.

As long as $\omega/T$ scaling is not violated \cite{abanov2003quantum,dell2007electrical,punk2016finite}, these overdamped excitations will strongly influence the temperature dependence of observables within the quantum critical region of Fig.~\ref{phasediag}. This region is delimited by the two crossover lines satisfying $k_BT \sim |h-h_{\rm QCP}|^{z\nu}$ with $z\nu=3/2$ according to our results. 
For instance, one can extract the critical contribution to the specific heat, which scales as $C \propto T^{(d - \theta)/z} = T^{\frac{4}{3} -\frac{2}{3} \theta}$. Here, $\theta$ is an exponent which describes hyperscaling violation. Usually hyperscaling violation occurs in systems with a critical Fermi surface, where the integral of the singular part of the free energy along the entire Fermi surface alters the thermodynamic properties \cite{PhysRevB.94.045133}.
In the context of the FFLO critical point discussed here, hyperscaling violation is not expected to occur for a sizeable velocity detuning $\delta v$, when the critical degrees of freedom live in the vicinity of isolated hot spots. Then, $\theta = 0$ and therefore  $C \propto T^{4/3}$.  This is similar to the SDW hot spots studied in Refs.~\cite{aavishkar,mandal2017scaling}. By contrast, for the case of vanishing velocity detuning to which our RG computation was restricted, the entire Fermi surface becomes hot. As a result, one expects  a hyperscaling violation exponent $\theta=1$ and therefore $C \propto T^{2/3}$. We emphasize again, however, that the hot spot theory (our field theoretical starting point) remains applicable in this limit as well: the infinite set of hot spot pairs decouple in the low energy limit, because electrons can only scatter with small momentum transfer tangential to the Fermi surface, similar to the Ising-nematic case. For this reason we are confident that our RG computation remains valid for finite velocity detunings as well, even though thermodynamic observables may depend strongly on the velocity detuning via the hyperscaling violation exponent $\theta$. 

From the low-energy form of  $\rho(k_0)$ of the hot quasiparticles one can also make a prediction for the temperature dependence of the NMR relaxation rate, $1/(T\,T_1) \propto T^{2/3}$ \cite{piazza2016fflo}. Note that for strong velocity detuning, the cold electrons give an additional constant contribution to $1/(T\, T_1)$ (Korringa law).

In organic superconductors, measurements of specific heat \cite{lortz2007calorimetric,exp_calorim_fflo} and NMR rates \cite{mayaffre2014evidence} within the putative quantum-critical region have been already taken. While one may see indication for non-Fermi liquid behaviour in the data (see Ref.\ \cite{piazza2016fflo}), quantitative statements and meaningful estimates on critical exponents cannot be made yet. A new round of data taking on a larger temperature interval might provide a conclusive insight.

\section{Competing orders}
\label{compordsec}

Non-Fermi liquid fixed points, where the critical correlations take a form similar to Eqs.\ (\ref{twopointscaling}), (\ref{Dtwopointscaling}),
arise in numerous physical contexts. As discussed above, in principle the zero-temperature form of the correlations manifests itself in a quantum-critical region at finite temperatures, see Fig.\ \ref{phasediag}. However, the critical scaling is often masked by a ``dome" of a competing, mostly superconducting order \cite{metlitski2010instabilities, lederer2015enhancement, mandal2016superconducting,mandal2017scaling}, at least for conventional critical points associated with the onset of broken symmetry \cite{metlitski2014non}. The FFLO-normal metal fixed point is different in this regard: since we deal with a phase transition towards superconductivity already, one can expect the fixed point to be ``naked". Other superconducting orders, e.g.\ of triplet type, may of course occur, but seem unlikely given the Fermi surface geometry of Fig.\ \ref{Fermisurface}, in accordance with  a recent Monte-Carlo study of a Hubbard model with spin imbalance \cite{gukelberger2016fulde}.
 
Going beyond these naive expectations, one may answer the question how competing instabilities are modified close to our new non-Fermi liquid fixed point systematically in the dimReg framework:  
Following the treatment of Ref.\ \cite{sur2015quasilocal}, we consider the insertion of a generic fermion bilinear into the critical action (\ref{critactionhigherdim}). In the spinor language, this term can be of two types: Either 
\begin{align}
\label{type1}
&\text{type}\  1: \quad 
 \lambda\int_{k^{d+1}} \bar{\Psi}_\alpha(k) A \Psi_\beta(k)  B_{\alpha\beta} \quad \text{or} \\ 
&\text{type}\  2: \quad 
\label{type2}
\lambda\int_{k^{d+1}}  \Psi^T_\alpha(k) A \Psi_\beta(-k)B_{\alpha\beta} + \text{h.c.} \ , 
\end{align}
where $A$ and $B$ are $2\times2$ hermitian matrices: $A$ acts in spinor space, while $B$ acts in spin-space. $\lambda$ is a real-valued scalar, which can be viewed as an external source field coupling to the respective order parameter. 

Restricting ourselves to instabilites where the bare vertex is momentum independent, a general vertex can be written as sum of such terms. As seen explicitly below, the quantum corrections do not mix at one-loop level, so it suffices to study the terms individually.

We aim to classify the quantum corrections $V$ to these operators at one-loop level. The corresponding diagrams are shown in Fig. \ref{general_vertex}. 

\begin{figure}
\centering
\includegraphics[width=\columnwidth]{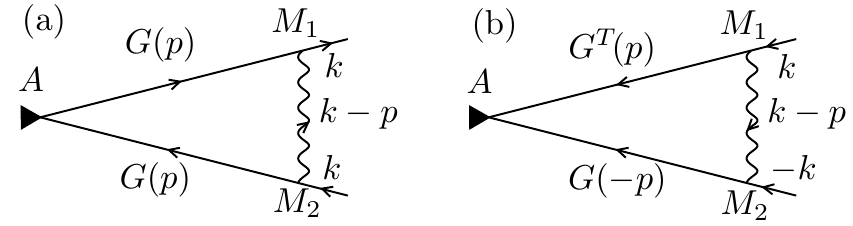}
\caption{Generic one-loop vertex correction in spinor space: (a) type 1 vertex (b) type 2 vertex.}
\label{general_vertex}
\end{figure}

In leading order in $\epsilon$, these diagrams  renormalize $\lambda$ as $\lambda \rightarrow \lambda (1+ u_{\lambda}g^{4/3}/\epsilon)$; for $u_{\lambda} >0 \ (<0)$, the instabilites are enhanced (suppressed). In RG formulation, the associated beta functions fulfill: 
\begin{align}
\beta_\lambda = \frac{d\lambda}{d\ln \mu} =  \lambda(-1 - \eta_\lambda),
\end{align}
with anomalous dimension $\eta_\lambda$. Proceeding as in the previous section, we find
\begin{align}
\label{anomlambda}
\eta_\lambda = \frac{2}{3} u_\lambda g^{4/3} =\bigg|_{g = g^\star} - \frac{u_\lambda}{u_g} \epsilon.  
\end{align}
To compute one-loop corrections $V$ to the fermion bilinears of Eqs.\ (\ref{type1}), (\ref{type2}), as a basis for the matrices $A,B$ we choose $\mathbbm{1}, \sigma_x, \sigma_y, \sigma_z$. 
The calculations are then fairly straightforward; technical details are presented in Appendix \ref{vertexApp}. Let us  sketch the results, starting with type 1 competing orders:
For $A = \mathbbm{1}$ or $A = \sigma_z$, the $\epsilon$-divergent vertex corrections are proportional to 
\begin{align} 
\label{antisymform} 
V \propto \int_{p_{d+1}} \vec{\Gamma}\cdot \vec{p} \cdot f(|\vec{p}|),
\end{align}  
where $\vec{\Gamma}$ is the vector of Gamma-matrices for the extra inserted dimensions (i.e., this vector has one entry in $d=3$), and $f$ is some function. In $d=2$, there are no extra dimensions, and (\ref{antisymform}) vanishes trivially. Indeed, type 1 corrections with diagonal spinor matrices $A$ correspond to superconducting instabilites; for these, the one-loop vertex correction is trivially absent as the diagram simply cannot be drawn. In higher dimensions, Eq.\ (\ref{antisymform}) also vanishes by antisymmetry. In particular, the FFLO boson-fermion vertex correction vanishes as already stated in Sec.\ \ref{diagsec}. 
Thus, superconducting vertices are not modified at the critical point at one-loop level. Of course, for pairing vertices one should also take into account momentum dependent form factors, but these should only render the vertex less RG-relevant. 

For $A= \sigma_x$, the corrections $V$ are shown to vanish as well, similar to the vertex corrections in the Ising-nematic case \cite{dalidovich2013perturbative}. Finally, for $A = \sigma_y$ the corrections vanish for $d=2$ only (by Cauchy's integral theorem). Near $d=5/2$, there are non-zero contributions; these lead to enhancement or suppression depending on the spin-matrix $B$. Writing out the vertex (\ref{type1}) in terms of ordinary fermions $\psi$, the results are summarized in Tab. \ref{tab1}.

\begin{table*}
\begin{ruledtabular}
\begin{center}
    \begin{tabular}{lll}
       \multicolumn{1}{l}{\textbf{Spin-matrix} $B$}   & \multicolumn{1}{l}{\textbf{Terms in action} } & 
   \multicolumn{1}{l}{\textbf{Anomalous dimension $\eta_\lambda$ }} \\ 
 	  \hline
 	\multicolumn{1}{ l }{$\mathbbm{1}$}
     &$\bar{\psi}_{\alpha}^R\psi_{\alpha}^R - \bar{\psi}_{\alpha}^L\psi_{1\alpha}^L$ & $\phantom{-}1.00 \epsilon$, \ enhanced \\ 
    % \hline
     \multicolumn{1}{ l }{$\sigma_x$}
     & $\bar{\psi}_{\uparrow}^R\psi_{\downarrow}^R+ \bar{\psi}_{\downarrow}^R\psi_{\uparrow}^R - \bar{\psi}_{\uparrow}^L\psi_{\downarrow}^L - \bar{\psi}_{\downarrow}^L\psi_{\uparrow}^L$ & $-1.00 \epsilon $, \ suppressed \\ 
  %   \hline
     \multicolumn{1}{ l }{$\sigma_y$}
    & $  i(\bar{\psi}_{\downarrow}^L\psi_{\uparrow}^L + \bar{\psi}_{\downarrow}^R\psi_{\uparrow}^R) -i(\bar{\psi}_{\uparrow}^R\psi_{\downarrow}^R + \bar{\psi}_{\uparrow}^L \psi_{\downarrow}^L)  $ & $\phantom{-}1.00\epsilon$, \ enhanced  \\
  %  \hline 
    \multicolumn{1}{ l }{$\sigma_z$}
    & $\bar{\psi}_{\uparrow}^R\psi_{\uparrow}^R - \bar{\psi}_{\downarrow}^R\psi_{\downarrow}^R - (\bar{\psi}_{\uparrow}^L\psi_{\uparrow}^L - \bar{\psi}_{\downarrow}^L \psi_{\downarrow}^L)$ &$ -1.00\epsilon $, \ suppressed   \\
 
    \end{tabular}
    \caption{Type 1 instabilities, of the form $\bar{\Psi}_\alpha \sigma_y \Psi_\beta B_{\alpha\beta}$, modified by the FFLO at one loop}
    \label{tab1}    
\end{center}
\end{ruledtabular}
\end{table*} 

Thus, the type 1 vertices influenced by FFLO fluctuations correspond to density interactions between fermions with the same sheet index, with relative phases locked in various ways. Let us go over to type 2 competing orders, as these are easier to interpret and quantitatively more important. In particular, they also pick up sizable corrections for $d=2$. For spinor-matrices $A= \mathbbm{1}, \sigma_z$, the quantum corrections vanish analogously to Eq.\ (\ref{antisymform}). For $A= \sigma_x, \sigma_y$, nontrivial corrections can arise. Evaluating all combinations  $\Psi^T_\alpha(k) A \Psi_\beta(-k)B_{\alpha\beta}$ is again straightforward and shown in Appendix \ref{vertexApp}; some combinations of $A,B$ vanish trivially due to anticommutation of fermion fields. The results are summarized in Tab.\ \ref{tab2}.

\begin{table*}
\begin{ruledtabular}
\begin{center}
    \begin{tabular}{llll}
\multicolumn{1}{l}{\textbf{Spinor} $A$}    
   &    \multicolumn{1}{l}{\textbf{Spin} $B$}   & \multicolumn{1}{l}{\textbf{Terms in action} } & 
   \multicolumn{1}{c}{\textbf{Anomalous dimension $\eta_\lambda$}} \\ 
 	  \hline
 	  $\sigma_x$ &
 	\multicolumn{1}{ l }{$\sigma_y$}
     &$2i(\bar{\psi}_{\downarrow}^L\psi_{\uparrow}^R - \bar{\psi}_{\uparrow}^L\psi_{\downarrow}^R) + \text{h.c.}$: \text{SDW in $y$-direction}& $-1.70\epsilon$, \ suppressed \\ 
    % \hline
    $\sigma_y$ &
     \multicolumn{1}{ l }{$\mathbbm{1}$}
     & $ 2i\bar{\psi}_{\alpha}^L\psi_{\alpha}^R + \text{h.c.} $: CDW at $2k_{F,\downarrow}$ or $2k_{F,\uparrow} $     & $\phantom{-}2.69\epsilon$, \ enhanced \\ 
  %   \hline
  $\sigma_y$ &
     \multicolumn{1}{ l }{$\sigma_x$}
    & $2i(\bar{\psi}_{\downarrow}^L\psi_{\uparrow}^R + \bar{\psi}_{\uparrow}^L\psi_{2\downarrow}^R)$ + h.c.: SDW in $x$-direction  & $-2.69\epsilon$, \ suppressed  \\
  %  \hline 
  $\sigma_y$ &
    \multicolumn{1}{ l }{$\sigma_z$}
    & $2i(\bar{\psi}_{\uparrow}^L\psi_{\uparrow}^R - \bar{\psi}_{\downarrow}^L\psi_{\downarrow}^R) $ + h.c.: SDW in $z$-direction  & $-2.69\epsilon$, \ suppressed   \\

    \end{tabular}
 \caption{Type 2 instabilities, of the form $\bar{\Psi}^T_\alpha A \Psi_\beta B_{\alpha\beta} + \text{h.c.}$, modified by the FFLO at one loop.}
    \label{tab2}    
\end{center}
\end{ruledtabular}
\end{table*}

As indicated in Tab.\ \ref{tab2}, competing orders that aquire a non-trivial one-loop correction from FFLO order correspond to the Spin Density-Wave (SDW) or Charge Density-Wave (CDW) channel. Only the latter order, with a wavevector peaked at $2k_{F,\downarrow}$ or $2k_{F,\uparrow}$, is enhanced.  Note that this order, which is referred to as $2k_F$-scattering in Ref.\ \cite{dalidovich2013perturbative}, is suppressed in the Ising-nematic case; the change in sign can be cross-checked by integrating out bosons and noting that the resulting effective four-fermion interaction has an opposite sign when decoupled in the $2k_F$-channel in the Ising-nematic case compared to the FFLO case. In summary, our analysis of instabilities indentifies the $2k_F$-CDW as the only serious competitor for FFLO criticality in $d=2$.

Of course, this dimReg computation can only predict how a tendency to order is enhanced, but not if there is an instability in the first place. A first indication that CDW order may indeed be important here can be obtained by straightforward evaluation of the corresponding vertex diagram with both fermions and bosons dressed by FFLO self-energies, which indeed shows a logarithmic divergence. To unambiguously answer the question which ordering tendency (FFLO or CDW) is more important, one would need to perform an RG analysis of an action which treats both orders on the same footing, e.g.\ similar to Ref.\ \cite{mandal2016superconducting}; we leave this task for future work.

\section{Conclusion and outlook}
\label{concsec}

In this work we have analyzed the quantum critical point between a FFLO superconductor and a normal metal phase in an anisotropic 2D system. Computing critical properties in a controlled expansion in $\epsilon = 5/2 - d$  dimensions we have found a non-Fermi liquid fixed point, characterized by a dynamical critical exponent $z=1+\epsilon$ and a correlation length exponent $\nu = 1 + \mathcal{O}(\epsilon^2)$ to leading order in $\varepsilon$. We derived the scaling forms of electronic and order-parameter correlations, and discussed possible physical manifestations.

One big advantage of the FFLO critical point compared to other non-Fermi liquid systems is that the scaling regime of the QCP is potentially accessible down to arbitrary low temperatures, if the quantum critical point is not masked by a competing order, such as superconductivity in heavy Fermion compounds or cuprate superconductors. In order to shed some light on this question
we also performed a general analysis of competing instabilities and found that charge density wave ordering is enhanced in the vicinity of the FFLO critical point. It is thus possible that the FFLO QCP is masked by a CDW phase in certain materials, depending on microscopic details. Extending our RG analysis to a situation where FFLO and CDW fluctuations are treated on equal footing would be an interesting problem for future study.
In a similar spirit, one could attempt an RG analysis of disorder \cite{PhysRevB.95.205106, *Mandal:2017rpj}, which is known to destroy the FFLO state in organic superconductors \cite{aslamazov1969influence}.

Our analytical derivation relies heavily on the approximation that the spin-up and spin-down Fermi surface branches have the same curvature respectively vanishing velocity detuning $\delta v \rightarrow 0$. While this parameter choice is physically grounded, treating the $\delta v \neq 0$ case e.g.\ numerically would be very interesting, potentially revealing a modification of the Fermi surface shape as in the SDW case \cite{Metlitski2010}. In addition, one could try to start from the opposite limit $\delta v \rightarrow \infty$.
A higher loop analysis of the problem would be desirable as well, but appears rather involved; alternatively, for $\delta v \neq 0$ one could apply the scheme with fixed co-dimension as shortly discussed, and see if it leads to similar results.

\acknowledgements
The authors acknowledge insightful discussions with D. Chowdhury, S. Huber, D.  Schimmel, and P. Strack. This work was supported by the German Excellence Initiative via the Nanosystems Initiative Munich (NIM).

\appendix 

%\titleformat{\subsection}
%  {\normalfont\itshape}{\thesubsection}{1em}{\centering}

\titleformat{\section}
			{\normalfont\bfseries}{Appendix \thesection:\!}{.5em}{\normalfont\bfseries}
			
\section{Mean field phase transition of a microscopic model}		

\label{meanfieldapp}	
			
To illustrate our field theoretic starting point, in this Appendix we recall the ordinary Ginzburg-Landau picture of the phase transition. Paraphrasing the treatment of Ref.\ \cite{piazza2016fflo}, we start from a microscopic model appropriate e.g.\ for the Bechgaard salt $(\text{TMTSF})_2\text{ClO}_2$ \cite{Jerome_fflo, lebed2010larkin}: we consider spinful fermions freely moving along chains oriented in $x$-direction, with a small interchain hopping parameter $t$. When these electrons are Zeeman-coupled to a magnetic field $h$, the free fermionic Hamiltonian reads 
\begin{align}
\label{Fermihamil}
H_0 &= \sum_{\alpha = \uparrow, \downarrow, \k} \xi_\alpha(\k) \psi^\dagger_{\alpha}(\k) \psi_{\alpha}(\k),  \\
\xi_{\alpha} &= k_x^2/2 - 2t\cos(k_y) - \mu - s_\alpha h, \ \{s_\uparrow, s_\downarrow\} = \{1,-1\}	\notag	, 
\end{align}			
where $\mu$ is the chemical potential, and we set the fermion mass and interchain distance to $1$. Plotting the Fermi surface with parameters $\mu = 3.3, t = 0.5, h = 1.0$ readily reproduces Fig.\ \ref{Fermisurface}.  

We now assume that the electrons interact with some short-range attractive interaction hamiltonian $H_{\text{int}}$ (e.g.\ mediated by phonons) as in Eq.\ \eqref{Hint} . Then, we introduce a functional integral representation of $H = H_0 + H_{\text{int}}$, resulting in a quantum action $S$ (see, e.g., Ref.\  \cite{Altland2010}). Decoupling the interaction term $H_{\text{int}}$ in the pairing channel  yields (we consider finite temperature $T$ for generality) 
\begin{align}
&\notag
S[\psi_\alpha, \bar{\psi}_\alpha, \Delta, \bar{\Delta}] = S_0[\psi_\alpha, \bar{\psi}_\alpha] + S_\text{int}[\psi_\alpha, \bar{\psi}_\alpha, \Delta, \bar{\Delta}], \\ \notag 
&S_\text{int} = \sum_{\omega_n,\q} g|\Delta(\omega_n,\q)|^2 - \\ &\frac{g}{\sqrt{\beta V}}  \sum_{\substack{\k,\q \\\nu_n, \omega_n}} \bar\Delta(\omega_n,\q) \psi_\downarrow(\omega_n - \nu_n, \q - \k)\psi_\uparrow(\nu_n,\k) + \ \text{h.c.},
\label{Appendixaction}
\end{align}
where $S_0$ is the bare fermionic action derived from Eq. \eqref{Fermihamil}, $\omega_n,\nu_n$ are bosonic and fermionic Matsubara frequencies, respectively, and $\beta$ is the inverse temperature. The subsequent mean-field analysis shows that the superconducting susceptibility is peaked at momenta $ \pm \textbf{Q}_{\text{FFLO}} = \left(k_{F,\uparrow}-k_{F,\downarrow}\right) \textbf{e}_x $, where $k_{F,\alpha}$ are the respective Fermi momenta of the two spin species. Consequently electrons interact with superconducting fluctuations $\Delta$ predominantly at so-called hotspots on the Fermi surface which are connected by $\textbf{Q}_\text{FFLO}$, found at $k_y = 0, k_x = \pm k_{F,\alpha}$. For this reason, within a low-energy theory sufficient for a universal RG analysis, we can expand the fermion fields as well as the fermion dispersions near these hotspots. In this manner, we introduce four low-energy fields $\psi_{\uparrow/\downarrow}^{L/R}$. Furthermore expanding $\Delta$ near $ \pm \textbf{Q}_{\text{FFLO}}$  readily yields action \eqref{veryfirstaction} in the limit $V\rightarrow \infty, T\rightarrow 0$ apart from different boson kinetic and mass terms, which automatically arise in the RG flow as discussed in the main text. 

A standard Landau-Ginzburg analysis of Eq.\ \eqref{Appendixaction}, which indicates a continuous phase transition, can be performed by integrating out the fermions.\footnote{This is dangerous for 2D fermionic systems, see e.g. Ch.\ 18 of Ref.\  \cite{Sachdev2011}; a proper analysis requires an RG procedure as presented in this paper. } This yields an effective bosonic action
\begin{align}
\label{bosonicaction}
S_{\Delta}[\Delta,\bar{\Delta}] = \sum_{\omega_n,\q} g|\Delta(\omega_n,\q)|^2 - \text{Tr}\, \text{ln} \, G^{-1}, 
\end{align}
where Tr denotes the trace in spin and energy-momentum space, and $G^{-1}$ is a matrix propagator: 
\begin{align}
\label{matprop}
&G^{-1}(\nu_n,\nu_n^\prime,\k,\k^\prime) = \\ & \notag \begin{pmatrix} \beta \delta_{\nu_n,\nu_n^\prime} \delta_{\k,\k^\prime} (i\nu_n - \xi_\uparrow(\k))  &   \sqrt{\beta/V}g\Delta(\nu_n - \nu_n^\prime, \k - \k^\prime) \\ \sqrt{\beta/V}g\bar\Delta(\nu_n - \nu_n^\prime, \k - \k^\prime)   &   \beta \delta_{\nu_n,\nu_n^\prime} \delta_{\k,\k^\prime} (i\nu_n + \xi_\downarrow(-\k)) \end{pmatrix}.
\end{align}		
To generally treat Eq.\ \eqref{bosonicaction} on mean field level, one would proceed by solving for the saddle point, $\delta/(\delta \Delta) S_{\Delta} \overset{!}{=} 0$, making an appropriate mean field ansatz for the (static) boson. The Larkin-Ovchinnikov ansatz, around which our dynamical boson in the main part is expanded, reads 
\begin{align}
\label{LOAnsatz}
\Delta_{\text{LO}}(\omega_n,\q) = \Delta_0 \delta_{\omega_n,0} \left(\delta_{\q, \textbf{Q}_{\text{FFLO}}} + \delta_{\q, -\textbf{Q}_{\text{FFLO}}}\right), 
\end{align}
where the amplitude $\Delta_0$ can be chosen real. However, a derivation of a closed-form saddle point equation ($\hat =$ mean field self-consistency equation) is difficult since it requires the inversion of Eq.\ (\ref{matprop}), which is hindered by the involved momentum dependence in Eq.\ (\ref{LOAnsatz}). To avoid this difficulty, one can plug in the ansatz \eqref{LOAnsatz} into $S_\Delta$ and expand in powers of $\Delta_0$ up to fourth order. Since the odd terms trivially vanish by symmetry, one obtains an effective Landau-Ginzburg functional \begin{align}
S_{\text{LG}}[\Delta_0] = m[h] \Delta_0^2 + a_4[h] \Delta_0^4,
\end{align} where we have indicated the magnetic field dependence explicitly. A strong indication for a continuous transition at mean field level is then given if (see, e.g., Ref.\ \cite{0953-8984-30-30-305604}) the boson mass $m$ can be tuned to zero for approprate $h$, while $a_4>0$. The second condition was shown to be true in Ref.\ \cite{piazza2016fflo} (see Appendix A within).
Let's focus on the first one here. As easily shown, the coefficient $m$ is given by
\begin{align}
\label{a2}
&m = 2g^2(1/g -  \Pi_0[h]), \\
& \Pi_0[h] = \sum_\k \frac{1 - n_F[\xi_{\uparrow} (\k)]- n_F[\xi_\downarrow(\textbf{Q}_{\text{FFLO}} - \k)]}{\xi_{\uparrow}(\k) + \xi_{\downarrow}(\textbf{Q}_{\text{FFLO}} - \k))}, 
\label{a2Pi}
\end{align}
where $n_F$ is the Fermi-distribution, and $\Pi_0$ the static inverse pair propagator respectively the boson self-energy. Evaluating Eq.\ \eqref{a2Pi} for general external boson momenta, one easily check's that it is indeed peaked at $\textbf{Q}_{\text{FFLO}}$ as claimed before.

 We limit ourselves to a numerical evaluation of $\Pi_0[h]$ in the limit $T\rightarrow 0$; a plot for generic parameters shown in Fig.\ \ref{PiFFLO}

\begin{figure}[H]
\centering
\includegraphics[width=\columnwidth]{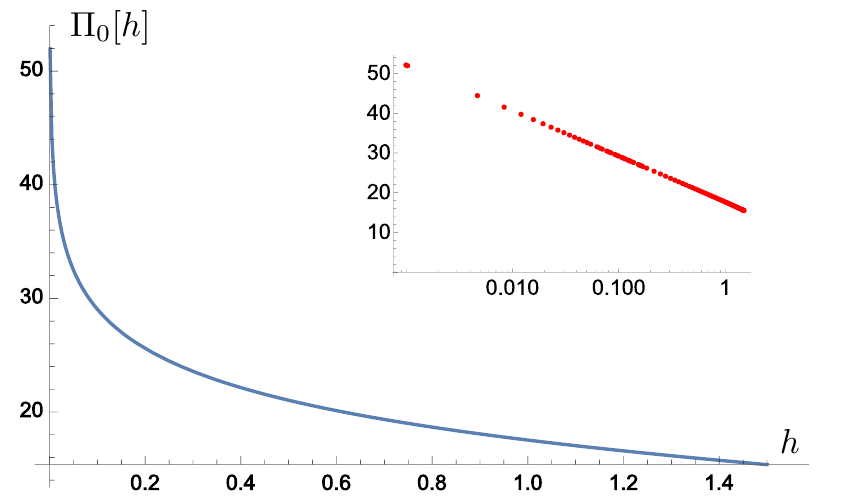}
\caption{$\Pi_0[h]$ numerically computed from Eq.\ \eqref{a2Pi}, with parameters $\mu = 3.3, t = 0.5$. The inset shows the same plot on a log-linear scale. }
\label{PiFFLO}
\end{figure} 

As clearly seen in Fig.\ \ref{PiFFLO}, $\Pi_0[h]$ diverges as $h \rightarrow 0$. In fact, this divergence is logarithmic, as pinpointed in the inset. This is in accordance with the analytical evaluation for the low-energy action in Appendix \ref{suplogApp} (where $\delta v \propto h$), and also with Ref.\ \cite{samokhin2006quantum}. Therefore, at any arbitrarily small value of the coupling $g$, there is a critical magnetic field $h_c \propto \exp(-1/g)$ where the mass term $m$ in Eq.\ \eqref{a2} changes sign, and the mean field phase transition between the normal metal and the FFLO superconductor occurs. Close to $h_c$, the field-dependence of the mass term scales as $m \propto (h-h_c)/h_c$, as claimed in the main text. 

The mean-field treatment presented above is fairly simplistic. First, it does not describe the phase transition between the FFLO and homogeneous superconductor -- to this aim one would have to make a homogeneous mean field ansatz as well, which we avoid since we are only interested in the QCP shown in Fig.\ \ref{phasediag}. One could also improve the mean field ansatz, say, by allowing for more complicated periodic functions that the $\cos(\textbf{Q}\cdot \x)$ LO-dependence, as done e.g.\ in Ref.\ \cite{mora2004nature}. We don't pursue this further since the mean-field treatment is not the focus of this work, and the general outcome that a mean-field transition exists and is continuous in 2D is generally agreed upon in the literature. 
\begin{comment}
{\color{red}NB: Es waer aber vllt allgemein interessant zu untersuchen wie sich der Fixpoint verandert wenn man um eine kompliziertere MeanField Loesung entwickelt (ein Boson feld mit mehreren componenten zulaesst?)}
\end{comment}

\section{Computation of self-energies}
\subsection{Boson self-energy}
\label{BoseSelfApp}

Here, we present the evaluation of the boson self-energy, given by Eq.\ (\ref{Bose selfenergy}): 
\begin{align}
\label{Bose selfenergyApp}
\Pi(k)  = -g^2\mu^{\epsilon} \int_{p^{d+1}}\sum_{\alpha \neq \alpha^\prime} \text{Tr}\left[G_\alpha (-p) M_1 G_{\alpha^\prime}(k-p) M_2\right].
\end{align}
To evaluate the trace, we use 
\begin{align}
\label{Bosetrace}
\text{Tr}[\sigma_i M_1 \sigma_j M_2] = 
\bordermatrix{
&  j=x  & j=y  & j=z \cr
 i=x & 1 & -i & 0 \cr
  i=y &i & 1 & 0 \cr
i=z & 0 & 0 & 0  }.  
\end{align}
In the limit $\delta v \rightarrow 0$ discussed in the main text, this leads to 
\begin{align}
\label{Pinotev}
\Pi(k)  = 2g^2 \mu^\epsilon \! \int\displaylimits_{p^{d+1}} \frac{(\delta_{-p} + i p_0)\cdot (\delta_{k-p} + i (k_0 - p_0))}{\left[\boldsymbol{P}^2 + (\delta_{-p})^2\right] \left[\left(\boldsymbol{K}-\boldsymbol{P}\right)^2 + (\delta_{k-p})^2\right]}. 
\end{align}
Changing to energy variables $x = \delta_{-p}, y = \delta_{k-p}$, with Jacobian $1/|2k_d|$, $\Pi(k)$ is rewritten as 
\begin{align}
\notag
\Pi(k) = &\frac{g^2\mu^{\epsilon}}{|k_d|} \int \frac{d p_0}{2\pi}  \frac{d \vec{p}}{(2\pi)^{d-2}}\frac{d x}{2\pi} \frac{d y}{2\pi} \frac{x+ip_0}{\P^2 + x^2} \times \\  &\frac{y+i(k_0-p_0)}{(\K - \P)^2 + y^2}, \quad \P = (p_0,\vec{p}). 
\end{align}
Note that the limit $\delta v \rightarrow 0$ is already required at this stage: for general velocity detuning, the Jacobian of the transformation to energy variables is more involved, and the integration range is nontrivial as well, obstructing further evaluation.  

Taking the elementary $x,y$ integrals (note that the log-divergent parts vanish by antisymmetry), results in
\begin{align}
\Pi(k) = \frac{g^2\mu^\epsilon}{4|k_d|} \int\frac{d p_0}{2\pi} \int \frac{d \vec{p}}{(2\pi)^{d-2}} \frac{(k_0 - p_0)p_0}{|\P||\K - \P|}  .
\end{align}

To proceed (the remaining steps are similar to Sec.\ A1 of Ref.\ \cite{dalidovich2013perturbative}), we introduce a Feynman parameter, using: 
\begin{align}
\frac{1}{\sqrt{A_1}\sqrt{A_2}} = \frac{1}{\pi} \int_0^1 \! dt \frac{1}{\sqrt{t(1-t)}} \frac{1}{tA_1 + (1-t)A_2}. 
\end{align}
Shifting $\P \rightarrow \P + (1-t)\K$, this gives: 
\begin{align}
\notag
\Pi(k) =& \frac{g^2\mu^\epsilon}{4\pi|k_d|} \int\frac{d p_0}{2\pi} \int \frac{d \vec{p}}{(2\pi)^{d-2}} \int_0^1 dt \frac{1}{\sqrt{t(1-t)}} \\ \times &\frac{(tk_0 - p_0)(p_0+ (1-t)k_0)}{\P^2 + t(1-t)\K^2}.
\end{align}
We note that the terms of the numerator linear in $p_0$ give no contribution by antisymmetry. After rescaling $\P\rightarrow\sqrt{t(1-t)}\P$, we are left with a $t$-integral of the form
\begin{align}
\int_0^1 dt (t(1-t))^{d/2-1} = \frac{\Gamma\!\left(\tfrac{d}{2}\right)^2}{\Gamma(d)}. 
\end{align}
Going to polar coordinates, the remaining integrals read: 
\begin{align}
\label{Pikdsub}
\notag
&\Pi(k) = \frac{g^2\mu^\epsilon}{|k_d|}\frac{\Gamma\left(\tfrac{d}{2}\right)^2}{\Gamma(d)} \cdot \frac{2^{1-d}}{\pi^{d/2} \Gamma\!\left(\tfrac{d}{2} - 1\right)} \int_{-\infty}^\infty \frac{dp_0}{2\pi} \\&  \int_0^\infty \! d|\vec{p}| \ |\vec{p}|^{d-3}  \left(\frac{p_0^2 - k_0^2}{|\vec{p}|^2 + p_0^2 + |\vec{k}|^2 + k_0^2} - \frac{p_0^2}{|\vec{p}|^2 + p_0^2}\right).
\end{align}
Here, we have also subtracted $\Pi(0,k_d)$ for UV regularization. 
As discussed in the main text, the residual momentum dependence of this subtraction can be seen as an artefact of the $\delta v \rightarrow 0$ limit, and is further discussed in Appendix 
\ref{suplogApp}. Formally, this subtraction can also be justified by referring to Veltman's formula (see e.g.\ Ref.\  \cite{kleinert2001critical}). 

 It is instructive to study the $|\vec{p}|$-integral as $d\!\rightarrow\!2$. In this limit, the extra dimensions vanish and the $|\vec{p}|$-integral should be absent. Indeed, as $d\!\rightarrow\!2$, the integral becomes IR log-divergent, and so comes from $\vec{p}=0$ only; the log-divergence is asymptotically canceled by the prefactor $\sim 1/\Gamma(d/2 -1)$.
The remaining integrations are staightforward, resulting in Eq.\ (\ref{PikfinalDL}) of the main text: 
\begin{align}
\label{PikfinalDLapp}
\Pi(k) &= \chi_d \frac{g^2\mu^{\epsilon}}{|k_d|} (d \cdot k_0^2+|\vec{k}|^2) \cdot (k_0^2 + |\vec{k}|^2)^{\tfrac{d-3}{2}}, \\ 
\chi_d &= \frac{\Gamma((1 -d)/2)}{2^{d+2}\pi^{(d+1)/2} }\cdot \frac{\Gamma(d/2)^2}{\Gamma(d)}. \notag 
\\ \chi_{5/2} &\simeq -0.0178. \notag 
\end{align} 
\subsection{Fermion self-energy}
\label{FermiSelfApp}

We continue with evaluation of the  fermion self-energy with external spin-index $\kappa$, starting from Eq.\ (\ref{Sigma_1}): 
\begin{align}
\label{Sigma_1App}
&\Sigma_1^\kappa(k) = g^2 \mu^{\epsilon} \int_{p^{d+1}} D(p) M_1 G_\beta (k+p) M_2 \sigma_y^{\kappa \beta} \sigma_y^{\beta \kappa}. 
\end{align}
The sums in spinor space can be performed using
\begin{align}
M_1 \sigma_i M_2 = \begin{pmatrix}
0 & 1 \\ 0 & 0 
\end{pmatrix}\cdot \begin{cases} \phantom{-}1 \quad i=x
\\
-i \quad i=y \\
\phantom{-}0 \quad i=z
\end{cases}.
\end{align}
In the spin-independent limit $\delta v \rightarrow 0$ this leads to 
\begin{align}
\label{sigma1DL}
&\Sigma_1 = \\ 
& \notag 
 \underbrace{\begin{pmatrix}  
0 & -i \\ 0 & 0 
\end{pmatrix}
\cdot g^2}_{\equiv c_1}\mu^{\epsilon} \int_{p^{d+1}} \frac{(k_0 + p_0)\cdot i  + \delta_{k+p}}{(\K + \P)^2 + (\delta_{k+p})^2} \cdot \frac{1}{p_d^2 - \Pi(p)}.
\end{align}
Inserting the boson self-energy, one can elementarily evaluate the $p_{d-1}, p_d$ integrals, resulting in
\begin{align}
\label{Sigma1interm}
&\Sigma_1 =  \frac{ic_1\mu^{\epsilon}}{3\sqrt{3}}\! \int \frac{d p_0}{2\pi}  \frac{d \vec{p}}{(2\pi)^{d-2}}  \frac{(k_0+p_0)}{|\K + \P|} \times \\ & \frac{1}{\mu^{\epsilon/3}\chi^{1/3}(d\cdot p_0^2+|\vec{p}|^2)^{1/3} \cdot (p_0^2 + |\vec{p}|^2)^{\tfrac{d-3}{6}} }, \quad \chi = -\chi_d g^2. 
\notag 
\end{align}
We apply a Feynman parametrization:
\begin{align}
\notag
\label{twoFeynman}
&\frac{1}{A_1^{\alpha_1} A_2^{\alpha_2} A_3^{\alpha_3}} = \frac{\Gamma(\alpha_1 + \alpha_2 + \alpha_3)}{\Gamma(\alpha_1)\Gamma(\alpha_2)\Gamma(\alpha_3)} \int_0^1 dt_1 \int_0^{1-t_1} dt_2 \\ & \frac{t_1^{\alpha_1 -1}t_2^{\alpha_2 -1} (1-t_1-t_2)^{\alpha_3 -1}}{(t_1 A_1 + t_2 A_2 + (1-t_1 -t_2) A_3)^{\alpha_1 + \alpha_2 + \alpha_3}} \ .
\end{align}
With $\alpha_1 = 1/2, \alpha_2=1/3, \alpha_3 = (d-3)/6$, Eq.\ (\ref{Sigma1interm}) is rewritten as
\begin{align}
\notag 
&\Sigma_1 = c_2 \int \frac{d p_0}{2\pi}  \frac{d \vec{p}}{(2\pi)^{d-2}} \int_0^{1} dt_1 \int _0^{1-t_1} \! dt_2 \\  &\frac{(k_0+p_0)\cdot t_1^{-\tfrac{1}{2}}t_2^{-\tfrac{2}{3}}(1-t_1-t_2)^{\tfrac{d-9}{6}}}{\left[t_1(\K + \P)^2 + t_2 (d p_0^2 + |\vec{p}|^2) + (1-t_1 -t_2)(p_0^2 + |\vec{p}|^2)\right]^{\tfrac{d+2}{6}}  } 
\notag \\
&c_2 = \frac{ic_1\mu^{2/3\epsilon}\cdot\Gamma({\tfrac{2+d}{6}})}{3\sqrt{3} \cdot \chi^{1/3} \cdot \Gamma(\tfrac{1}{2})\Gamma(\tfrac{1}{3})\Gamma(\tfrac{d-3}{6})}.
\label{sigmafeyn}
\end{align}
Strictly speaking, the Feynman parametrization of Eq.\ (\ref{sigmafeyn}) is only well-defined for $d>3$, as the $t_2$-integral is otherwise divergent. We will circumvent this problem by evaluating the $t_2$-integral for general $d>3$ below (after the momentum integrals), and then analytically continue the result to $d<3$; the divergence at $d=3$ will cancel against the term $\Gamma((d-3)/6)$ contained in the factor $c_2$. As there certainly is a strip of convergence of the original integral (\ref{Sigma1interm}), and we also recover the $d=2$ result of Ref. \cite{piazza2016fflo}, this procedure should be legitimate.
To proceed, in Eq.\ (\ref{sigmafeyn}) we shift 
\begin{align}
\label{feynershift}
p_0 \rightarrow p_0 + \frac{-t_1}{1+t_2(d-1)} k_0, \quad \vec{p} \rightarrow \vec{p} -t_1\vec{k}.
\end{align}
Disregarding the linear terms in $p_0$ which vanish by antisymmetry, we then obtain: 
\begin{align}
\label{d1d2def}
&\Sigma_1 = c_2 \int \frac{d p_0}{2\pi}  \frac{d \vec{p}}{(2\pi)^{d-2}} \int_0^{1} dt_1 \int _0^{1-t_1} \! dt_2  \\ \notag &\frac{(1+\tfrac{-t_1}{1+t_2(d-1)})k_0\cdot t_1^{-\tfrac{1}{2}}t_2^{-\tfrac{2}{3}}(1-t_1-t_2)^{\tfrac{d-9}{6}}}{\left(d_1 k_0^2 + d_2 p_0^2 + |\vec{p}|^2 + |\vec{k}|^2 t_1(1-t_1) \right)^{\tfrac{d+2}{6}}  } \\ & d_1 = t_1 - \frac{t_1^2}{1+(d-1)t_2} \ , \quad d_2 = 1+(d-1)t_2. \notag
\end{align}
For $2\leq d<5/2$, the momentum integrals can be straightforwardly evaluated by going to polar coordinates, yielding
\begin{align}
\label{sigma1almostdone}
\notag
\Sigma_1 =  &k_0  \frac{c_2\cdot 2^{1-d}\cdot\Gamma(\tfrac{4}{3} - \tfrac{d}{3})\cdot\Gamma\left(\frac{\epsilon}{3}\right)}{\pi^{(d-1)/2}\cdot \Gamma(\tfrac{2+d}{6})\cdot\Gamma\left(\frac{1}{2} + \frac{\epsilon}{3} \right)}  \int_0^{1} dt_1 \int _0^{1-t_1} \! dt_2  \\ &\frac{(1-\frac{t_1}{1+(d-1)t_2}) t_1^{-\tfrac{1}{2}}t_2^{-\tfrac{2}{3}}(1-t_1-t_2)^{\tfrac{d-9}{6}}}{\sqrt{1+ (d-1) t_2}} \times \notag   \\ &\left(\frac{\mu^2}{d_1 k_0^2 + t_1(1-t_1)|\vec{k}|^2}\right)^{\epsilon/3}.
\end{align}
Following the procedure described below Eq.\ (\ref{sigmafeyn}), let us evaluate the $t_2$-integral for $d>3$. With an eye for the final limit $\epsilon \rightarrow 0$, we still set the last dimensionless prefactor in Eq.\ (\ref{sigma1almostdone}) equal to one, which should be fine as this is a perfectly regular function in $t_1,t_2$. We have also checked this numerically on a simplified integral. Furthermore, note that in $d=2$, we can extract a factor of $|k_0|^{-1/3}$ from the integral, and obtain a self-energy  $\propto |k_0|^{2/3}$ as found in Ref.\ \cite{piazza2016fflo}.

Evaluation of the $t_2$-integral yields, without the other prefactors, the fairly involved expression: 
\begin{widetext}
\begin{align} \notag
\tilde F(d) = &
- \Gamma
   \left(\tfrac{d-3}{6}\right) \times (1-t_1)^{\frac{d-7}{6}} \Gamma \left(\tfrac{1}{3}\right) \bigg\{ \bigg[\left(d \left(t_1^2-3
   t_1+3\right)-t_1 (t_1+3)\right) 
    \ \!_2F_1\left(\tfrac{1}{3},\tfrac{1}{2};\tfrac{d-1}{6};(d-1)
   (t_1-1)\right)  \\ &+(4-d) t_1 \,
   _2F_1\left(-\tfrac{1}{2},\tfrac{1}{3};\tfrac{d-1}{6};(d-1)
   (t_1-1)\right)\bigg]\bigg\}/\left[3 \sqrt{t_1} (d (t_1-1)-t_1)
   \Gamma \left(\tfrac{d-1}{6}\right)\right].
\end{align}
\end{widetext}
where $_2F_1$ is the hypergeometric function. $\tilde{F}(d)$ is divergent for $d \!\searrow \!3$ due to the prefactor $\Gamma((d-3)/6)$, but this factor cancels against the same factor contained in the overall prefactor $c_2$ [c.f.\ (\ref{sigmafeyn})]. The remainder
$
F(d) \equiv \tilde{F}(d)/ \Gamma((d-3)/6)$
is a well-behaved function. Its numerical integration leads to  
\begin{align}
\int_0^1 dt_1 F\left(\tfrac{5}{2}\right) \simeq 1.166
\end{align}
Collecting all prefactors, and expanding the Gammafunctions from Eq.\  (\ref{sigma1almostdone}) in $\epsilon$, one obtains 
\begin{align} \notag 
\Sigma_1 &= \frac{u_g g^{4/3}}{\epsilon} \cdot \begin{pmatrix} 
0 & -i \\ 0 & 0 \end{pmatrix}  \cdot (-ik_0) + \text{finite terms}, \\  u_g &= -0.0813  .  
\end{align}
Evaluation of $\Sigma_2$ given in Eq.\ (\ref{Sigma_2}) proceeds analougously. In total, one arrives at Eq.\ (\ref{fermiselfmaintext}) of the main text: 
\begin{align}
\label{fermiselfapp}
\Sigma(k) &= \Sigma_1(k) + \Sigma_2(k) \\& =
 \frac{u_g g^{4/3}}{\epsilon}  \cdot \sigma_y \cdot (-ik_0) + \text{finite terms}. \notag
\end{align}

\section{Computation of vertex corrections for competing instabilities}
\label{vertexApp}

In this Appendix, we compute the anomalous dimensions of possible competing orders, which are summarized in Tables \ref{tab1}, \ref{tab2}.

\subsection{Type 1 orders}
 As in the main text, we start with type 1 orders, computing one-loop corrections $V$ to the fermion bilinear of Eq.\ (\ref{type1}). Fixing the signs with Wick's theorem, in the limit $\delta v \rightarrow 0$ where the Green's functions become spin-independent, they have the general form 
 \begin{align}
\label{Vtype1}
&V = \lambda\int_{k^{d+1}} \bar{\Psi}_\gamma(k) \Omega(k)  \Psi_{\delta}(k) \cdot \left(\sigma_y^{\beta\delta}\sigma_y^{\gamma\alpha} B_{\alpha\beta}\right),  \\ 
&\Omega(k) =
 \Omega^1(k) + \Omega^2(k) \notag\\ &   \Omega^1(k) = g^2 \mu^{\epsilon} \int_{p^{d+1}}\!M_1G(p) A G(p) M_2 D(k-p) \notag  \\  &\Omega^2(k) = \Omega^1\left[M_1 \!\leftrightarrow\! M_2\right].
\end{align}
Let's fix $A = \mathbbm{1}$ and compute $\Omega^1$. The sums in spinor space are determined from 
\begin{align}
\label{type1firstspinor}
M_1\sigma_i \mathbbm{1} \sigma_j M_2 = \begin{pmatrix} 0 & 1 \\ 0 & 0 \end{pmatrix} \cdot
\bordermatrix{
&  j=x  & j=y  & j=z \cr
 i=x & 0 & 0 & -1 \cr
  i=y &0 & 0 & i \cr
i=z & 1 & -i & 0  }. 
\end{align}
Since $G \propto -\boldsymbol{\Gamma}\cdot \P + \sigma_x \delta_p$ and we take the Gamma-matrices $\vec{\Gamma}$ in the extra dimensions to be proportional to $\sigma_z$ [c.f. Eq.\ (\ref{Gammavec})], it immediately follows that $\Omega^1$ is of the form 
\begin{align} 
\label{antisymformapp} 
\Omega^1 \propto \int_{p_{d+1}}\! (\vec{\Gamma}\cdot \vec{p}) \cdot f(|\vec{p}|),
\end{align} 
where $f$ is some function. This expression vanishes as discussed in the main text below Eq.\ (\ref{antisymform}). The same conclusion holds for $A = \sigma_z$. For $A = \sigma_x$, using
\begin{align}
M_1 \sigma_i \sigma_x \sigma_j M_2 = \begin{pmatrix} 0 & 1 \\ 0 & 0 \end{pmatrix} \!\cdot
\!\bordermatrix{
&  j=x  & j=y  & j=z \cr
 i=x & 1 & -i & 0 \cr
  i=y &-i & -1 & 0 \cr
i=z & 0 & 0 & -1 },  
\end{align}
we obtain
\begin{align}
\label{typesigmaxgone}
\Omega^1 =- g^2 \mu^{\epsilon}\! \begin{pmatrix}
0 & 1 \\ 0 & 0 \end{pmatrix}
\int_{p^{d+1}} \frac{-\P^2 + \delta_p^2 + 2ip_0\delta_p}{\left(\delta_p^2 + \P^2\right)^2} D(k-p).  
\end{align}
This expression has the same form as the vertex correction in the Ising-nematic case \cite{dalidovich2013perturbative}. Since the boson propagator $D$ is independent of $p_{d-1}$, after shifting $p_{d-1} \rightarrow \delta_p$ Eq.\ (\ref{typesigmaxgone}) vanishes due to the identity
\begin{align}
\int dx \frac{ x^2 - a^2}{\left(x^2 + a^2\right)^2} = 0. 
\end{align}
Last, we consider $A = \sigma_y$. 
Using
\begin{align}
\label{type1sigmaydspinor}
M_1\sigma_i \sigma_y \sigma_j M_2 = \begin{pmatrix} 0 & -i \\ 0 & 0 \end{pmatrix}\! \cdot\!
\bordermatrix{
&  j=x  & j=y  & j=z \cr
 i=x & -1 & i & 0 \cr
  i=y &i & 1 & 0 \cr
i=z & 0 & 0 & -1}, 
\end{align}
we find  
\begin{align}
\label{Omegatype1sigmay}
&\Omega^1 = \\ \notag 
& -
\underbrace{ g^2 \!
\begin{pmatrix} 0 & -i \\ 0 & 0 \end{pmatrix}}_{\equiv c_1} \mu^{\epsilon} \int_{p^{d+1}} \frac{p_0^2 - p^2 - \delta_p^2 - 2i\delta_pp_0}{(\P^2 + \delta_p^2)^2} D(k-p). 
\end{align}
Performing the $p_{d-1}$-integral (by shifting $p_{d-1} \rightarrow \delta_p$), we get 
\begin{align}
\label{Omstep1}
\Omega = c_1 \mu^{\epsilon}  \int \frac{dp_0}{2\pi} \frac{d\vec{p}}{(2\pi)^{d-2}} \frac{dp_d}{2\pi} \frac{|\vec{p}|^2}{2|\P|^3} D(k-p).
\end{align}
Note that for $d=2$, Eq. (\ref{Omegatype1sigmay}) vanishes by Cauchy's integral theorem, which can be seen by reducing the fraction; accordingly, the integrand in Eq.\ (\ref{Omstep1}) is proportional to the external momenta $|\vec{p}|^2$. To further evaluate Eq. (\ref{Omstep1}), we focus on  $k = (k_0, 0 \hdots)$, which is sufficient in leading order in $\epsilon$. Shifting $p_0 \rightarrow p_0 + k_0$ for convenience and performing the $p_d$-integral gives 
\begin{align}
&\Omega^1 = \frac{c_1\mu^{2\epsilon/3} }{3\sqrt{3}\chi^{1/3}}  \int \frac{dp_0}{2\pi} \frac{d\vec{p}}{(2\pi)^{d-2}} \\ \notag & \frac{p^2}{((p_0+k_0)^2+p^2)^{3/2}\cdot (dp_0^2 + p^2)^{1/3}\cdot (p_0^2+p^2)^{(d-3)/6}}, 
\end{align}
with $\chi \simeq 0.0178 g^2$. Applying the Feynman-parametrization (\ref{twoFeynman}), with $\alpha_1 = 3/2, \alpha_2 = 1/3, \alpha_3 = (d-3)/6$, we obtain 
\begin{widetext}
 
\begin{align}
\label{shortwidetext}
\Omega^{1}_{\alpha\beta} =  &\underbrace{\frac{c_1 }{3\sqrt{3}\chi^{1/3}}\cdot \frac{\Gamma\left(\frac{d+8}{6}\right)}{\Gamma(\tfrac{3}{2})\Gamma(\tfrac{1}{3})\Gamma\left(\frac{d-3}{6}\right)}}_{\equiv c_2} \mu^{2\epsilon/3} \int \frac{dp_0}{2\pi} \frac{d\vec{p}}{(2\pi)^{d-2}} \int_0^1 dt_1 \int_0^{1-t_1} dt_2 \ t_1^{1/2}t_2^{-2/3}(1-t_1-t_2)^{\tfrac{d-9}{6}} \times \\ &\frac{p^2}{\left[(p_0 + k_0)^2 + p^2)t_1 + (dp_0^2+ p^2)t_2 + (p_0^2+p^2)(1-t_1-t_2)\right]^{\frac{d+8}{6}}}, \notag
\end{align}

\end{widetext}
where we follow the same logic as in the evaluation of Eq.\ (\ref{sigmafeyn})f. Shifting $p_0 \rightarrow p_0 - \frac{t_1}{(d-1)t_2 +1}k_0$ and going to polar coordinates yields

\begin{align} \notag 
&\Omega^1 =  \frac{c_2\mu^{\epsilon} 2^{3-d}}{\pi^{d/2-1}\Gamma\!\left(\tfrac{d}{2}-1\right)} \!\int_{-\infty}^{\infty} \!\frac{dp_0}{2\pi} \int_0^\infty \! d|\vec{p}|  \int_0^1 \! dt_1 \int_0^{1-t_1} \! dt_2 \\ &t_1^{1/2} t_2^{-2/3} (1-t_1-t_2)^{\frac{d-9}{6}}  \frac{|\vec{p}|^{d-1}}{(d_1 k_0^2 + d_2 p_0^2 + |\vec{p}|^2)^{\frac{d+8}{6}}}, 
\end{align}
where $d_{1/2}$ were defined in Eq.\ (\ref{d1d2def}).
Performing the $|\vec{p}|, p_0$-integrals is then straightfoward and results in: 
\begin{align}
\label{omegasigmaxalmost}
&\Omega^1 = \frac{c_2\Gamma\left(\tfrac{d}{2}\right)\Gamma\left(\tfrac{\epsilon}{3}\right)}{4\sqrt{\pi}\Gamma\left(\tfrac{d+8}{6}\right)} \int_0^1 dt_1 \int_0^{1-t_1} dt_2 \\ \notag & \frac{t_1^{1/2} t_2^{2/3} (1-t_1 - t_2)^{\frac{d-9}{6}}}{\sqrt{d_2} \Gamma\left(\frac{d-3}{6}\right) }\left(\frac{\mu^2}{d_1 k_0^2}\right)^{\epsilon/3}.
\end{align}
Approximating the last expression in  parentheses in 
Eq.\ (\ref{omegasigmaxalmost}) by 1, the $t_2$-integral can be evaluated analytically for $d>3$; the divergence as $d\rightarrow3$ cancels against the factor $\Gamma\left[(d-3)/6\right]$ contained in $c_2$, c.f.\ Eq.\ (\ref{shortwidetext}). Then, the $t_1$-integral can be computed numerically for $d=5/2$, yielding 
\begin{align}
\Omega^1 \simeq \begin{pmatrix} 0 & -i \\ 0 & 0 \end{pmatrix} \cdot  0.081 \frac{g^{4/3}}{\epsilon}.
\end{align}
$\Omega^2$ [c.f.\ Eq.\ (\ref{Vtype1})]  is evaluated in the same vein, and in total we obtain
\begin{align}
\Omega = \sigma_y \cdot 0.081 \frac{g^{4/3}}{\epsilon}.
\end{align}
Now, we need to evaluate the factor involving the spin-matrix $B$ in Eq.\ (\ref{Vtype1}), which yields: 
\begin{align}
\left(\sigma_y^{\beta\delta}\sigma_y^{\gamma\alpha} B_{\alpha\beta}\right) = \begin{cases} \phantom{-} B_{\gamma\delta}, \quad B = \mathbbm{1}, \sigma_y \\ 
- B_{\gamma\delta}, \quad B = \sigma_x, \sigma_z . 
\end{cases}
\end{align}
Alltogether, the quantum correction $V$ therefore reads 
\begin{align}
\notag
V &= \lambda\frac{u_{\lambda}g^{4/3}}{\epsilon}\! \int_{k^{d+1}} \!\bar{\Psi}_{\gamma}(k) \sigma_y \Psi_\delta(k) \!\times \! \begin{cases} \phantom{-} B_{\gamma\delta}, \;  B = \mathbbm{1}, \sigma_y \\ 
- B_{\gamma\delta}, \; B = \sigma_x, \sigma_z
\end{cases} \\  
u_\lambda &= 0.081. 
\end{align}
Using Eq.\ (\ref{anomlambda}), this readily yields Tab.\ \ref{tab1}.

\subsection{Type 2 orders}

We proceed with type 2 orders, computing corrections $V$ to the fermion bilinear of Eq.\ (\ref{type2}). Analogous to the previous case, they are of the form 

 \begin{align}
\label{Vtype2}
&V = \lambda\int_{k^{d+1}} \Psi^T_\gamma(k) \Omega(k)  \Psi_{\delta}(-k) \cdot \left(\sigma_y^{\beta\delta}\sigma_y^{\alpha\gamma} B_{\alpha\beta}\right) + \text{h.c.},  \\ 
&\Omega(k) =
 \Omega^1(k) + \Omega^2(k) \notag\\ &   \Omega^1(k) = g^2 \mu^{\epsilon} \int_{p^{d+1}}\!M_1G^T(p) A G(-p) M_2 D(k-p) \notag  \\  &\Omega^2(k) = \Omega^1\left[M_1 \!\leftrightarrow\! M_2\right].
\label{Omegastype2}
\end{align}

For $A = \mathbbm{1}, \sigma_z$, $\Omega^1$ vanishes as in the previous case. For $A = \sigma_x$, the required product in spinor space reads

\begin{align}
\label{type2first}
M_1\sigma_i^T\! \sigma_x \sigma_j M_2 = \begin{pmatrix} 0 & 1 \\ 0 & 0 \end{pmatrix} \! \cdot \!
\bordermatrix{
&  j=x  & j=y  & j=z \cr
 i=x & 1 & -i & 0 \cr
  i=y &i & 1 & 0 \cr
i=z & 0 & 0 & -1},
\end{align}
resulting in

\begin{align}
\notag
\Omega^1 = \underbrace{\begin{pmatrix} 0 & 1 \\ 0 & 0 \end{pmatrix} \cdot g^2}_{\equiv c_1} \mu^\epsilon \! \int_{p^{d+1}} \! & \frac{p_0^2-p^2 + ip_0 \delta_{-p} + ip_0 \delta_p - \delta_{p}\delta_{-p}}{\left(\P^2 + \delta_p^2\right)\left(\P^2 + \delta_{-p}^2\right)} \\ & \times D(k-p).   
\label{Omega1firsttype2} 
\end{align}
To evaluate this expression, we restrict ourselves to $k = (0, \hdots, k_d$). Then, the linear terms in $p_0$ vanish by antisymmetry. Taking the $p_{d-1}$-integral  results in

\begin{align}
\notag
\Omega^1 = &\frac{c_1 \mu^\epsilon}{2} \int \frac{dp_0}{2\pi} \frac{d\vec{p}}{(2\pi)^{(d-2)}} \frac{dp_d}{2\pi}   \frac{p_0^2}{|\P|(p_d^4 + \P^2)} 
\\  \times &\frac{1}{(k_d + p_d)^2 + \frac{\chi\mu^\epsilon}{| k_d + p_d |} \left( d p_0^2 + \vec{p}^2 \right) \left( p_0^2 + \vec{p}^2\right)^{\frac{d-3}{2}}},  \label{Omfirsttype2pd1} 
\end{align}
with $\chi \simeq 0.0178 g^2$. To evaluate Eq.\ (\ref{Omfirsttype2pd1}), we shift $p_d \rightarrow p_d - k_d$. Then, following Ref.\ \cite{dalidovich2013perturbative}, we may approximately disregard the $p_d$-dependence of the fermion part in leading order in $g$ (and hence in leading order in $\epsilon$).  We can then perform the $p_d$-integral, yielding
\begin{align}
\Omega^1 = \left(\frac{\mu}{k_d^2}\right)^{2/3 \epsilon} \frac{c_1}{\chi^{1/3}3\sqrt{3}} &\int \frac{dp_0}{2\pi} \frac{d\vec{p}}{(2\pi)^{(d-2)}} \\ &\frac{p_0^2}{(\P^2)^{d/6}(\P^2 + 1)\left( d p_0^2 + \vec{p}^2 \right)^{1/3} }, 
\notag
\end{align}
where we have also rescaled $\P \rightarrow \P/k_d^2$. In leading order in $\epsilon$, the first factor can be approximated by $1$.  The Feynman-parametrization (\ref{twoFeynman}) with  
$\alpha_1 = d/6, \alpha_2  = 1, \alpha_3 = 1/3$ then leads to 
\begin{widetext}

\begin{align}
\Omega^1 = &\frac{c_1}{6\sqrt{3}} \frac{\Gamma\left(\tfrac{4}{3} + \tfrac{d}{6}\right)}{\Gamma\left(\tfrac{1}{3}\right)\Gamma\left(\tfrac{d}{6}\right)}
 \int \frac{dp_0}{2\pi} \frac{d\vec{p}}{(2\pi)^{(d-2)}} \int_0^1 dt_1 \int_0^{1-t_1} dt_2 \ t_1^{d/6-1} (1-t_1 -t_2)^{-2/3} \\  & \times
\frac{p_0^2}{\left(t_1 (p_0^2 + |\vec{p}|^2) + t_2 (p_0^2 + |\vec{p}|^2 +1) + (1-t_1 - t_2)(d p_0^2 + |\vec{p}|^2)\right)^{4/3 + d/6} }.
\notag
\end{align}

\end{widetext}
Changing to polar coordinates, the integrals over $p_0,\vec{p}$ and $ t_2$ are straightforwardly computed. The remaining $t_1$-integral can be evaluated numerically for $d=5/2$. Performing the same steps for $\Omega^2$ [c.f.\ Eq.\ (\ref{Omegastype2})], in total one obtains,
in leading order in $\epsilon$: 
\begin{align}
\Omega = \Omega^1 + \Omega^2 \simeq  \sigma_x \cdot 0.138 \frac{g^{4/3}}{\epsilon}
\end{align}
Let us now consider $A = \sigma_y$. Using 
\begin{align}
\label{spinor2Asigmay}
M_1\sigma_i^T \!\sigma_y \sigma_j M_2 = \begin{pmatrix} 0 & -i \\ 0 & 0 \end{pmatrix} \!\cdot\!
\bordermatrix{
&  j=x  & j=y  & j=z \cr
 i=x & -1 & i & 0 \cr
  i=y &-i & -1 & 0 \cr
i=z & 0 & 0 & -1},  
\end{align}
we obtain  
\begin{align}
\notag
\Omega^1 = \begin{pmatrix} 0 & -i \\ 0 & 0 \end{pmatrix} \!\cdot\! (- g^2 \mu^\epsilon)
 \!\int_{p^{d+1}} \!&\frac{\P^2 + ip_0 \delta_{-p} + ip_0 \delta_p - \delta_{p}\delta_{-p}}{\left(\P^2 + \delta_p^2\right)\left(\P^2 + \delta_{-p}^2\right)} \\
  & \times D(k-p). 
  \label{2Asigmayfirst}
\end{align}
The computations proceed largely analogous to the previous case of $A = \sigma_x$; in total, we obtain
\begin{align}
\Omega \simeq -\sigma_y \cdot 0.219 \frac{g^{4/3}}{\epsilon}.
\end{align}

To proceed, we need to evaluate the factor involving the spin-matrix $B$ in Eq.\ (\ref{Vtype2}), which yields: 
\begin{align}
\left(\sigma_y^{\beta\delta}\sigma_y^{\alpha\gamma} B_{\alpha\beta}\right) = \begin{cases} - B_{\gamma\delta}, \quad B = \mathbbm{1}, \sigma_y \\ 
\phantom{-} B_{\gamma\delta}, \quad B = \sigma_x, \sigma_z . 
\end{cases}
\end{align}

Before denoting which contributions are enhanced and which are suppressed, we notice that some products under consideration vanish trivially: 
\begin{align}
\Psi_\alpha^T  \sigma_x \Psi_\beta B_{\alpha\beta}&= (\bar{\psi}_{\alpha}^L \psi_{\beta}^R - \bar{\psi}_{\beta}^L \psi_{\alpha}^R) B_{\alpha \beta} = 0 \\  & \qquad \qquad \qquad \ \text{for} \ B = \mathbbm{1}, \sigma_x, \sigma_z,
\notag \\
\Psi_\alpha^T  \sigma_y \Psi_\beta \sigma_y^{\alpha\beta}  &= i (\bar{\psi}_{\beta}^L \psi_{\alpha}^R + \bar{\psi}_{\alpha}^L \psi_{\beta}^R) \sigma_y^{\alpha\beta} = 0. 
\end{align}
Alltogether, the non-vanishing quantum corrections are, for $A = \sigma_x$:  

\begin{align}
V &= \lambda \frac{u_{\lambda}g^{4/3}}{\epsilon}\! \int_{k^{d+1}} \!\Psi^T_{\gamma}(k) \sigma_x \Psi_\delta(-k) \sigma_y^{\gamma\delta}, \quad 
u_\lambda = -0.138. 
\label{Vfinal1}
\end{align}
For $A = \sigma_y$: 
\begin{align}
\notag
V &= \lambda \frac{u_{\lambda}g^{4/3}}{\epsilon}\! \int_{k^{d+1}} \!\Psi^T_{\gamma}(k) \sigma_y \Psi_\delta(-k)\! \times\! \begin{cases} -B_{\gamma\delta}, \; B = \mathbbm{1}\\ 
\phantom{-}B_{\gamma\delta}, \; B = \sigma_x, \sigma_z
\end{cases} \\  
u_\lambda &= -0.219. 
\label{Vfinal2}
\end{align}
Using Eq.\ (\ref{anomlambda}), Eqs.\ (\ref{Vfinal1}), (\ref{Vfinal2}) readily yield Tab.\ \ref{tab2}.

\section{Superconducting logarithm}
\label{suplogApp}

To clarify the role of the limit $\delta v 
\rightarrow 0$ applied in this paper, it is instructive to reevaluate the boson self-energy of Eq.\ (\ref{Bose selfenergy})  for $\delta v \neq 0$ and $d=2$. Eq.\ (\ref{Bose selfenergy}) then reads, up to constant prefactors:

\begin{align}\notag 
\Pi(k) &\propto \int_{p^{2+1}}\sum_{\alpha \neq \alpha^\prime} \frac{1}{ip_0 - \delta_{-p}^\alpha} \frac{1}{i(k_0-p_0) - \delta_{k-p}^{\alpha^\prime}}, \\ \delta^\alpha_p &= v_\alpha p_x + p_y^2. 
\end{align}
Performing the integral over $p_0$ with help of Cauchy's theorem gives: 
\begin{align}
\label{appbstarting point}
\Pi(k)  \propto \int dp_x dp_y &\frac{1}{ik_0 - \delta_{-p}^\alpha - \delta^{\alpha^\prime}_{k-p}}  \\ \times&\left(\theta(\delta_{-p}^\alpha)\theta(\delta_{k-p}^{\alpha^\prime}) - \theta(-\delta^\alpha_{-p})\theta(-\delta_{k-p}^{\alpha^\prime})\right),   
\notag 
\end{align} 
where $\alpha^\prime \neq \alpha$. 
We introduce momentum cutoffs in the 2 directions $p_x$ and $p_y$, $\Lambda_x$ and $\Lambda_y$, with $\Lambda_x \simeq \Lambda_y^2$. Then, the $p_x$ integral in  (\ref{appbstarting point}) gives: 
\begin{align} 
&\Pi(k) = \\ \notag &\frac{1}{v_+}\!\int\!dp_y\bigg\{  \log\left(ik_0 + v_+ p_x - p_y^2 - v_{\alpha^\prime}k_x - (k_y - p_y)^2 \right)\! \bigg|_{-\Lambda_x}^{\text{Mi}}  \\ &- \log\left(ik_0 + v_+ p_x - p_y^2 - v_{\alpha^\prime}k_x -(k_y - p_y)^2 \right)\bigg|_{\text{Ma}}^{\Lambda_x}\bigg\} \ , \notag \\ \notag
&v_+ = v_\alpha + v_\alpha^\prime, \\
&\text{Mi}/\text{Ma} = \text{min}/\text{max}\left(\frac{p_y^2}{v_\alpha}, k_x + \frac{(k_y - p_y)^2}{v_\alpha^\prime}\right) . 
\end{align}
Inserting the boundaries $\pm\Lambda_x$ yields terms of the form $2\log(v_+\Lambda_x) + i\pi\text{sign}(k_0) + \mathcal{O}\left[(k,p)/\Lambda_x\right]$. These constant terms vanish once we subract $\lim_{k_0 \rightarrow 0} \Pi(k_0, 0)$, which is legitimate when working at the critical point. By noticing that, if $\text{Ma} = p_y^2/v_\alpha$ in some integration region $R_1$, then $\text{Mi} = p_y^2/v_\alpha$ in $\mathbb{R}  \setminus R_1$, we can recast the remainder in the following form: 
\begin{align}
 \notag
&\Pi(k) \sim \!\int\!dp_y\bigg\{  \log\left(ik_0 + vp_y^2 - v_{\alpha^\prime}k_x - (k_y - p_y)^2 \right) \\  &+ \log\left(ik_0 + v^{-1}(k_y-p_y)^2 + v_{\alpha}k_x - p_y^2 \right)\bigg\}  \label{twologs}, 
\end{align}
where $v= v_{\alpha^\prime}/v_\alpha$, and w.l.o.g.\ we assume $v>1$. The remaining integral can be straightforwardly evaluated; inserting the boundaries $\pm \Lambda_y$ yields a long expression, which is of the schematic form 
\begin{align}\notag
\label{Pikschema}
&\Pi(k) \propto \frac{1}{v-1}\sqrt{\left(ik_0-v_{\alpha^\prime}k_x\right)(v-1) - k_y^2v} \\  &+ \frac{1}{|v^{-1} -1|}\sqrt{\left(ik_0 + k_x v_{\alpha}\right)(v^{-1} -1) - k_y^2 v^{-1}} + \Pi_{\text{div}}, \notag 
\\ &\Pi_{\text{div}} \simeq \Lambda_y + \log\left( (\delta v)^2  + \frac{|k_y|}{\Lambda_y}\right). 
\end{align}
The first two terms of Eq.\ (\ref{Pikschema}) reproduce the result of Ref.\  \cite{piazza2016fflo}.  For these terms, the limit $\delta v\rightarrow 0$, which is equivalent to $v \rightarrow 1$, can be taken, and results in a standard damping term; see also Appendix 
\ref{codimApp}. Let's now consider $\Pi_{\text{div}}$, the divergent part of Eq.\ (\ref{Pikschema}):
 For the first summand, $\Lambda_y$, the limit $\Lambda_y \rightarrow \infty$ corresponds to a pure UV divergence, which effectively arises from expansion of the fermion dispersion in the low-energy action (\ref{veryfirstaction}). If higher  order terms in the dispersion are taken into account, this UV singularity is absent, as numerically demonstrated in Ref.\ \cite{piazza2016fflo}; we can therefore disregard this term. The second term is finite for $\delta v \neq 0$. In a fully realistic model of the FFLO transition, this condition is always fulfilled: Increasing the magnetic field leads to increasing $\delta v$, and the phase transition takes place when 
$g_0 - \Pi(\delta v,0)$ vanishes (on mean field level); here, $g_0$ is the strength of the original four-fermion interaction. This happens at a small but nonzero value $\delta v  = \delta v_c$. Thus, for $\delta v\simeq \delta v_c$, and $k_y \ll \Lambda_y$, i.e.\ when taking the limit $\Lambda_y \rightarrow \infty$ first, $\Pi_{\text{div}}$ is just a finite mass term, which can be dropped when performing computations at the critical point. The remainder is regular in $\delta v$, and one can take the limit $\delta v\rightarrow 0$ to simplify the computation.

On the other hand, in the dimReg computation we have to take the limit $\delta v \rightarrow 0 $ first, [c.f.\ Eq.\ (\ref{Pinotev})],    and are therefore left with the IR divergent quantity $\log(|k_y|/\Lambda_y)$, a standard ``BSC logarithm''. To correct for this unphysical way of taking the limits, one must subtract $\Pi_{\text{div}}(k_y)$, as effectively done in Eq.\ (\ref{Pikdsub}).

\section{Dimensional regularization with fixed co-dimension}

\label{codimApp}

In this work, we have performed a dimReg procedure by increasing the co-dimension of the Fermi surface. An alternative approach, shortly discussed in this Appendix, is to keep the co-dimension fixed, following Refs.\ \cite{chakravarty1995transverse, lee2017recent,mandal2015ultraviolet}. That is, in the higher-dimensional action the kinetic term for the fermions is modified to 
\begin{align}
\label{newex}
\int_{k^{d+1}} &\bar{\Psi}_\alpha(k) \left(-ik_0 \sigma_y + i(v_\alpha k_1 + \boldsymbol{K}^2)\sigma_x\right) \Psi_\alpha(k), \notag  \\ &\boldsymbol{K} = {(k_2, \cdots, k_d)},
\end{align}
with all other terms in the action unchanged. The leading terms in the action are then scale-invariant under 
\begin{align}
\label{fixedco}
k_{0} &= \frac{k^\prime_{0}}{b},\ \ k_1 = \frac{k_1^\prime}{b},\   \boldsymbol{K} = \frac{\boldsymbol{K}^\prime}{\sqrt{b}}, \\ \Psi &= b^{\tfrac{d}{4} + \tfrac{5}{4}}\Psi^\prime(k^\prime), \ \Delta(k) = b^{\tfrac{d}{4} + \tfrac{5}{4}} \Delta^\prime(k^\prime).
\end{align}
With this scaling, the interaction term becomes marginal in $d=3$, s.t.\ one can expand in $\epsilon = 3-d$.
In this scheme, evaluation of the Bose self-energy is very similar to the 2D-case sketched in Appendix \ref{suplogApp}. It can be performed in the general case $\delta v \neq 0$ by employing the trivial reshuffling decribed above Eq.\ (\ref{twologs}). 
Taking all momentum cutoffs to infinity, and subtracting $\Pi(0)$ for regularization (which works for $\delta v \neq 0$, see Appendix \ref{suplogApp}), one arrives at 
\begin{align}\notag
\label{BoseselffixedCo}
&\Pi(k) = \sum_{\alpha \neq \alpha^\prime} \frac{\beta_d}{\cos(d\pi/2)} \bigg\{ \left(\frac{ik_0 - k_1 v_{\alpha^\prime} - \K^2 \tfrac{v}{v-1}}{v-1}\right)^{\tfrac{d-1}{2}}  \\ &+ \left(\frac{ik_0 + k_1 v_{\alpha} - \K^2 \tfrac{v^{-1}}{v^{-1}-1}}{v^{-1}-1}\right)^{\tfrac{d-1}{2}} \bigg\},  
\end{align}
where $\beta_d > 0$ is a $d$-dependent factor of order 1. For $d\rightarrow 3$, $\Pi$ is $\epsilon$-divergent due to the term $\cos(d\pi/2)$. To gain analytical control, one can again expand in $\delta v$, which leads to 
\begin{align}
\notag 
\lim_{\delta v \rightarrow 0} \Pi(k) = 
& \frac{\beta_d}{\cos(d\pi/2)|\delta v|^{d-1}}\bigg\{ 4\cos\left(\tfrac{(d-1)\pi}{2}\right) |\K|^{d-1} \\  & - 4 \tfrac{(d-1)}{2} \cos(d\pi/2)|k_0 \delta v| |\K|^{d-3} \bigg\}.
\label{Pitexp}
\end{align}
In $d=2$, the prefactor of the term $\propto |\K|^{d-1}$ vanishes, and the remainder is the damping term of Ref.\ \cite{piazza2016fflo}, and regular as $\delta v \rightarrow 0$. However, for $2<d<3$, the first term does not vanish, and $\Pi(k)$ is divergent as $\delta v \rightarrow 0$. This can be seen as an instance of UV/IR mixing \cite{mandal2015ultraviolet}: As discussed in the main text [see Fig.\ \ref{singularlimitpatches2}], for $\delta v \rightarrow 0$ spin-up and spin-down Fermi sheets have the same curvature. As a result, any spin-up electron with momentum $\k_1$ on the Fermi surface can scatter against a spin-down electron with momentum $-\k_1$. However, if the Fermi surface is one-dimensional, the final states of this scattering event must have momenta $\pm \k_2 \simeq \k_1$; otherwise, the tangent vectors to the Fermi surface differ strongly, and the phase space for the scattering is negligible. By contrast, for a Fermi surface with dimension greater than one, all points of the Fermi surface share a mutual tangent vector. Therefore,  low-energy scattering events entangle the full Fermi surface, and the hot spot theory breaks down, as signaled by the $\delta v \rightarrow 0$ divergence of Eq.\ (\ref{Pitexp}).

\section{Magnetic susceptibility}
\label{suscapp}

In this Appendix, we shortly present the evaluation of the magnetic susceptibility close to criticality. We limit ourselves to evaluation of the functional behaviour (up to a constant prefactor). 

If the contribution of the fermions is neglected (or, phrased differently, they have been integrated out on one-loop level), the free energy on the normal metal side reads, for $d=2$: 
\begin{align}
\notag 
 F_\Delta &= - \ln [Z_\Delta]  \\ & =- \ln \left[ \int \mathcal{D}(\Delta,\bar{\Delta}) \exp( - \int d^3 k D^{-1}(k)|\Delta(k)|^2)\right] \notag \\&\propto \ln\left[{\det(D^{-1})}\right] = \int d^3k \left[\ln(D^{-1}(k))\right] .  
\end{align} 
Therefore, the fluctuation contribution to the magnetic susceptibility is given by \cite{larkin2008fluctuation,samokhin2006quantum}
\begin{align}
&\chi_\Delta \propto - \frac{\partial^2 F_\Delta}{\partial h^2} \propto -\frac{\partial^2 F_\Delta}{\partial m^2}  \propto  \int d^3 k \frac{-1}{\left(m + k_y^2 + \frac{\alpha|k_0|}{|k_y|}\right)^2},
\end{align}
where we reintroduced the mass term ($m>0$) into  the 2D boson propagator [see Eqs.\ (\ref{twopointscalingbosonssum}),(\ref{Dtwopointscaling})], and used that $m$ is proportional to the reduced magnetic field, $m = m_0 \frac{h-h_c}{h_c}$; $m_0$ and  $\alpha \propto g^2$ are constants. 
Easy integration yields 
\begin{align}
\chi_\Delta  \propto \Lambda_x \ln\left(\frac{\Lambda_y^2}{\frac{h-h_c}{h_c} \cdot m_0}\right), 
\end{align}
where $\Lambda_x, \Lambda_y$ are UV cutoffs in the $x,y$ directions (of order of Fermi energies). Normalizing $\chi_\Delta$ with the Pauli spin susceptibility in the normal state $\chi_P$ as in Ref.\ \cite{samokhin2006quantum}, and fixing the prefactors, on can conclude
\begin{align}
\frac{\chi_\Delta}{\chi_P} \simeq \frac{\Delta_0}{E_F} \ln \left(\frac{h_c}{h-h_c}\right),
\end{align}
where $\Delta_0$ is the BCS-gap and $E_F$ the Fermi energy.

\bibliographystyle{apsrev4-1}

\bibliography{../FFLO.bib}

\end{document}